\documentclass[reprint,amsmath,amssymb,braket,aps,prb,longbibliography,footinbib]{revtex4-2} 
\usepackage{graphicx,color,braket,amsmath,dcolumn,bm,float}
\usepackage[colorlinks=true,linkcolor=blue]{hyperref}
\usepackage{dblfnote}

\newcommand{\corra}[1]{\textcolor{black}{#1}}
\newcommand{\corrb}[1]{\textcolor{black}{#1}}

\begin{document}
\title{
Unconventional Quantization of 2D Plasmons in Cavities Formed by Gate Slots}

\author{Ilia Moiseenko$^{1}$, Zhanna Devizorova$^{1}$, Olga Polischuk$^{2}$, Viacheslav Muravev$^{3}$, Dmitry Svintsov$^{1}$}
\affiliation{$^{1}$Center for Photonics and 2d Materials, Moscow Institute of Physics and Technology, Dolgoprudny 141700, Russia}
\affiliation{$^{2}$Kotelnikov Institute of Radio Engineering and Electronics (Saratov Branch), Russian Academy of Sciences, Saratov 410019, Russia}
\affiliation{$^{3}$Osipyan Institute of Solid State Physics RAS, Chernogolovka, 142432 Moscow, Russia}
\email{svintcov.da@mipt.ru}

\begin{abstract}
We demonstrate that the slot between parallel metal gates placed above two-dimensional electron system (2DES) forms a plasmonic cavity with unconventional mode quantization. The resonant plasmon modes are excited when the slot width $L$ and the plasmon wavelength $\lambda$ satisfy the condition $L = \lambda/8 +n \times \lambda/2$, where $n=0, 1, 2 \ldots$. The lowest resonance occurs at a surprisingly small cavity size, specifically one eighth of the plasmon wavelength, which contrasts with the conventional half-wavelength Fabry-Perot cavities in optics. This unique quantization rule arises from a non-trivial phase shift of $-\pi/4$ acquired by the 2D plasmon upon reflection from the edge of the gate. The slot plasmon modes exhibit weak decay into the gated 2DES region, with the decay rate being proportional to the square root of the separation between the gate and the 2DES. Absorption cross-section by such slots reaches $\sim 50$ \% of the fundamental dipole limit without any matching strategies, and is facilitated by field enhancement at the metal edges.
\end{abstract}
\maketitle

Plasmonics, the study of collective electron oscillations in metallic and semiconductor nanostructures, has revolutionized the fields ranging from nanophotonics to sensing and energy harvesting by enabling light manipulation at subwavelength scales~\cite{Maier:2007}. At the heart of this discipline lie surface plasmons, the electromagnetic waves coupled to free electron oscillations at metal-dielectric interfaces, offering unprecedented control over light-matter interactions~\cite{Barnes:2003}. In turn, surface 2D plasmons propagating in two-dimensional electron systems feature ultrastrong confinement of electromagnetic energy accompanied by {\it in situ} tunability by the gate electrodes~\cite{Stern1967, Woessner2015, Ni_limits_plasmonics, Ham:2012}.

While the propagation of plasmons along flat surfaces is well studied for a range of materials~\cite{Ritchie_Plasmons,Stern1967,chaplik1972possible,Ryzhii2007a}, their behavior at non-uniformities is rich and complex. The latter mediate the conversion between surface waves and free-space photons~\cite{Woessner_phase_shifter,Alonso-Gonzalez_antenna_launching,Abajo_Limits}, and scatter the surface waves to other directions. Refraction of surface waves even at the simplest line discontinuity does not follow the conventional Fresnel's laws~\cite{Maradudin1990,agranovich1981diffraction,Pincemin1994,Ditlbacher2002}, though the derivation of alternatives faces considerable challenges. These stem from the two-dimensional propagation of surface waves and three-dimensional extent of their fields. An exact solution for surface wave scattering by the wedge dates back to Malyuzhinets~\cite{malyuzhinets1958excitation}, who generalized the Sommerfeld's diffraction problem~\cite{Sommerfeld:1896} to the finite surface impedance. The advent of high-quality two dimensional electron systems (2DES) and 2D materials stimulated further search for the plasmonic refraction laws~\cite{Moreno_phase}. These were eventually derived for the linear 2d junctions and 2DES terminations~\cite{Khavasi_pi4_phase,Alymov_Refraction}, applied to the analysis of complex cavities~\cite{Fogler_1d_junction,Semenenko_scattering}, and confirmed experimentally~\cite{Jadidi2015a,Tamagnone2018}. Another basic element of 2d plasmonic structures, the edge of a metal gate above 2DES, is still lacking its refraction laws. The gate edges are ubiquitous in electrically-tunable 2d plasmonics, and understanding their scattering properties is vital for design of electromagnetic detectors~\cite{Olbrich2016,Otsuji:2011}, sources~\cite{Kachorovskii_gated_instability,Boubanga-Tombet2020}, sensors~\cite{Xu_Sensor,Metasurface_sensor2}, and nonlinear photonic devices~\cite{AlonsoCalafell2021}. \corra{Ref.~\cite{Siaber2019} summarized the available analytical approaches~\cite{Sydoruk_gate_edge,Aizin_finite} to plasmon reflection at gate edges and found their considerable disagreement with simulations. In particular, the plane wave matching method~\cite{Aizin_finite} predicted a trivial reflection phase (zero or $\pi$), contrasting the simulations. An approximate analytical variational approach~\cite{Sydoruk_gate_edge,Siaber2019} predicted the reflection phase of $2{\rm arctan}(2/\pi) \approx65^\circ$, disagreeing with simulations by $\sim 40$ \%.  }

\begin{figure}
\center{\includegraphics[width=0.9\linewidth]{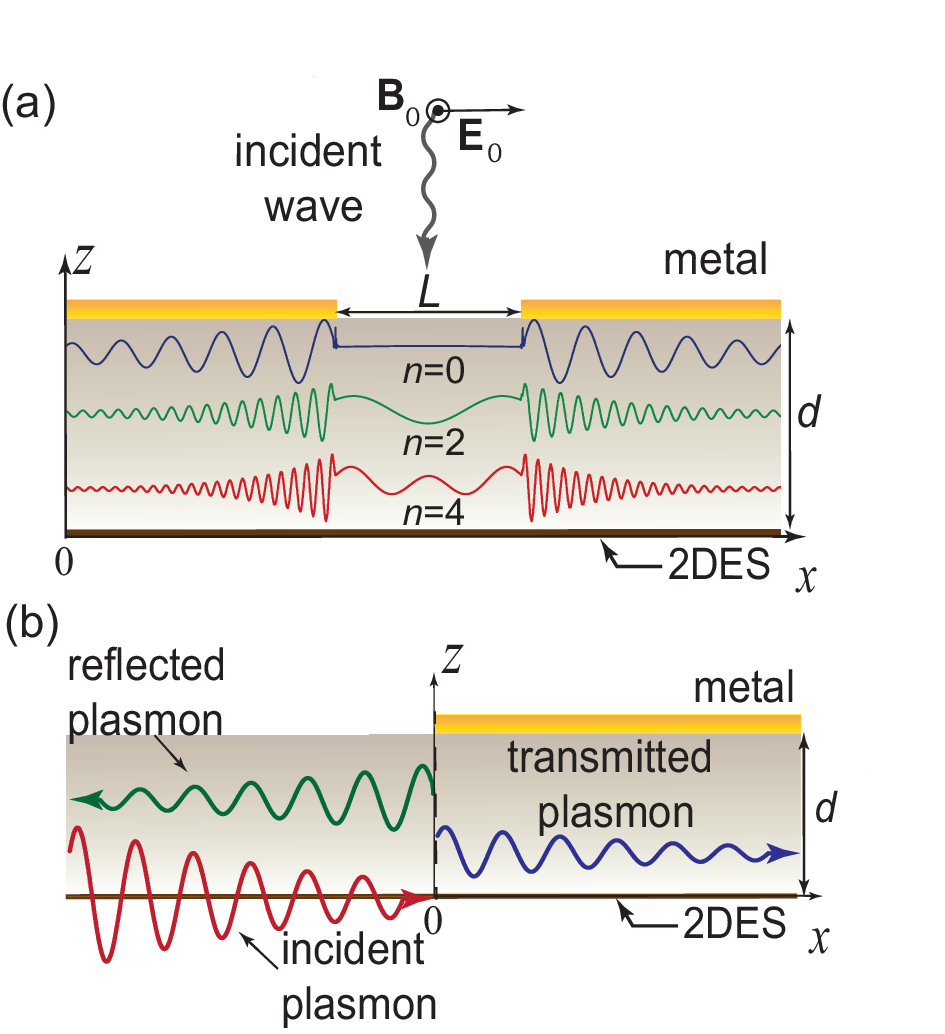}}
\caption{(a) Schematic illustration of a slot cavity for 2D plasmons. Electric field distributions for the three lowest bright cavity modes, excited by an incident plane wave, are shown with blue, green and red lines. (b) Schematic illustration of plasmon reflection at a single gate edge above 2DES. \corrb{Dashed line shows the reflection plane.}} 
\label{structure}
\end{figure}

In this Letter, we derive the quantitative reflection laws for 2D plasmons at the edges of metal gates, and use these laws for analysis of plasmonic cavities formed by gate slots. We find that the plasmon gains a non-trivial phase shift of $-\pi/4$ upon reflection from the gate edge, while the absolute reflectance is close to unity for small gate-channel separation. As a result, the gate slot above 2DES (shown in Fig.~\ref{structure} a) acts as a high-quality cavity for unscreened plasmons, with eigenmodes satisfying an unconventional quantization rule $L = \lambda/8 + n \times \lambda/2$, where $n=0, 1, 2 \ldots$ and $\lambda$ is the plasmon wavelength. Interestingly, the fundamental resonance in the slot is excited provided $L=\lambda/8$, which differs significantly from the familiar expression for frequency in an optical Fabry-Perot cavity, where $L=\lambda/2$.  We further show that the plasmon resonances have a finite linewidth even in a non-dissipative 2DES, which arises from leakage into propagating gated plasmon modes and radiative coupling. The electromagnetic absorption cross-section for the slot plasmon mode is very large, approaching the 'dipole limit' of $2\lambda_0/\pi$, where $\lambda_0$ is the free-space wavelength. 


The plasmon reflectance at the gate edge, $r$, is evaluated for the setup shown in Fig.~\ref{structure} b. The 2DES occupies the plane $z=0$ and is covered by a semi-infinite thin perfectly conducting gate located at $z=d$, $x>0$. The TM-polarized 2D plasmon with with wave vector $q_u$, frequency $\omega$, and electric field ${E}_{\rm inc} = E_0(z) {{e}^{i{{q}_{u}}x-i\omega t}} \theta \left( -x \right)$, is incident from the ungated region $x<0$ and causes the sought-after scattered fields $E_{\rm scat}$.

The governing equation for $E_{\rm scat}$ is obtained in several steps which we detail in Appendix A. First, the total field $E$ is linked to external sources $E_{\rm ext}$, the surface currents in 2DES $j_s$ (units of A/m), and the surface currents in the gate $j_g$ using the fundamental solution of the wave equation. This link is most convenient after the Fourier-transform with respect to $x$-coordinate, $q$ being the Fourier variable. Second, the surface currents in 2DES are eliminated~\cite{Moiseenko_partly_gated, Zabolotnykh_proximity, Mikhailov1998} using the Ohm's law $j_s(q)=\sigma(\omega) E(q,z=0)$, where $\sigma = \sigma'+ i \sigma''$ is the complex surface conductivity of 2DES per unit area with units of Ohm$^{-1}$, single and double primes distinguish between real and imaginary parts. The dispersion function $\sigma(\omega)$ can be arbitrary, and can be independently determined from kinetic models~\cite{Falkovsky2007a} or spectroscopic experiments~\cite{dahl2007magneto}. Focusing on electrodynamic aspects, we refrain from microscopic modeling of $\sigma$, and use it as a free parameter. Third, the fundamental solution of wave equation is written in the gate plane $z=d$ and links the field $E(q) = {E}_{\rm inc}(q) + E_{\rm scat}(q)$ and surface currents in the gate $j_g(q)$. The result reads as:
\begin{equation}
\left[ \frac{i{{E}_{0}}}{q-{{q}_{u}}}+{{E}_{\rm scat}}\left( q \right) \right]\frac{\varepsilon_u \left( q \right) }{{{\varepsilon }_{g}}\left( q \right)}\frac{k_0}{\kappa(q)} + \frac{i}{2} j_g(q) Z_0=0,
\end{equation}
where $Z_0$ is the free-space impedance equal to $4\pi/c$ in Gaussian units and 377 Ohm in SI units, $\varepsilon_u(q)$ and $\varepsilon_g(q)$ are the dielectric functions of the ungated and gated parts of the 2DES:
\begin{gather}
\varepsilon_u =1+i\eta \frac{\kappa(q)}{{{k}_{0}}},\\
{{\varepsilon }_{g}}=1+i\eta \frac{\kappa(q)}{{{k}_{0}}}\left( 1-{{e}^{-2\kappa(q)d}} \right),
\end{gather}
$k_0$ is the wave-vector of the electromagnetic wave propagating in free space, $\kappa(q) = \sqrt{q^2-k_0^2}$ is the decay constant of the electromagnetic field in the $z$-direction, and $\eta = \sigma Z_0/2$ is the dimensionless 2DES conductivity normalized by the free space impedance. 

The solution of such class of equations is achieved with Wiener-Hopf technique~\cite{Noble1958MethodsBO}. It implies collecting the functions analytic in the upper (+) and lower (-) half-planes of the complex $q$-variable in the left- and right-hand sides, respectively, and equating both sides to zero. Such manipulation requires the multiplicative splitting of all emerging functions into the lower ($-$) and upper ($+$) analytic parts, $f(q) = f_+(q) f_-(q)$
. The ''plus'' and ''minus'' functions $f_{\pm}$ are obtained from the original function $f$ with Cauchy theorem:
\begin{equation}
{f_{\pm }}\left( q \right)=\exp \left\{ \pm \frac{1}{2\pi i}\int\limits_{-\infty }^{+\infty }{\frac{\ln f \left( u \right)du}{u-\left( q\pm i\delta  \right)}} \right\}.
\end{equation} 

Introducing the shorthand notation $M(q)=k_0 \varepsilon_u \left( q \right) /{{\varepsilon }_{g}\left( q \right)}\kappa(q)$, performing the splitting, and equating the parts analytic in the upper and lower complex half-planes to zero, we get the solution for the scattered field:
\begin{equation}
\label{eq-scat-solution}
{{E}_{\rm scat}}\left( q \right) = -\frac{i{{E}_0}}{q-{q_u}}\left[ 1-\frac{{M}_{+}\left( q_u \right)}{{{M}_{+}}\left( q \right)} \right].
\end{equation}
The field $E_{\rm scat}$ does not simply represent the reflected plane wave, and its Fourier spectrum is relatively broad. This is a consequence of highly non-local electrodynamics in two dimensions. The plasma wave incident at the edge excites not only the reflected wave, but also non-trivial evanescent fields. Despite this complexity, the amplitude of reflected wave $r E_0$ is readily singled out from the scattered field (\ref{eq-scat-solution}). It is given by the residue of ${{E}_{\rm scat}}\left( q \right)$ at the pole $q=-q_u$ timed by $-i$~\cite{Kay_reactance_discontinuity,Alymov_Refraction}.
After several straightforward transformations (Appendix B), we arrive at
\begin{equation}
\label{eq-reflectance-full}
{r}=-\frac{i}{2{q_u}}{{\left[ \frac{{{\varepsilon }_{u+}}\left( {{q}_{u}} \right)}{{{\varepsilon }_{g+}}\left( {{q}_{u}} \right)} \right]}^{2}}\sqrt{\frac{{{q}_{u}}/{{k}_{0}}-1}{{{q}_{u}}/{{k}_{0}}+1}}\frac{e^{ -2 \kappa(q_u)d }}{{{\left. \partial \varepsilon_u /\partial q \right|}_{q=-{q_u}}}}.
\end{equation}

\begin{figure}[ht]
\center{\includegraphics[width=\linewidth]{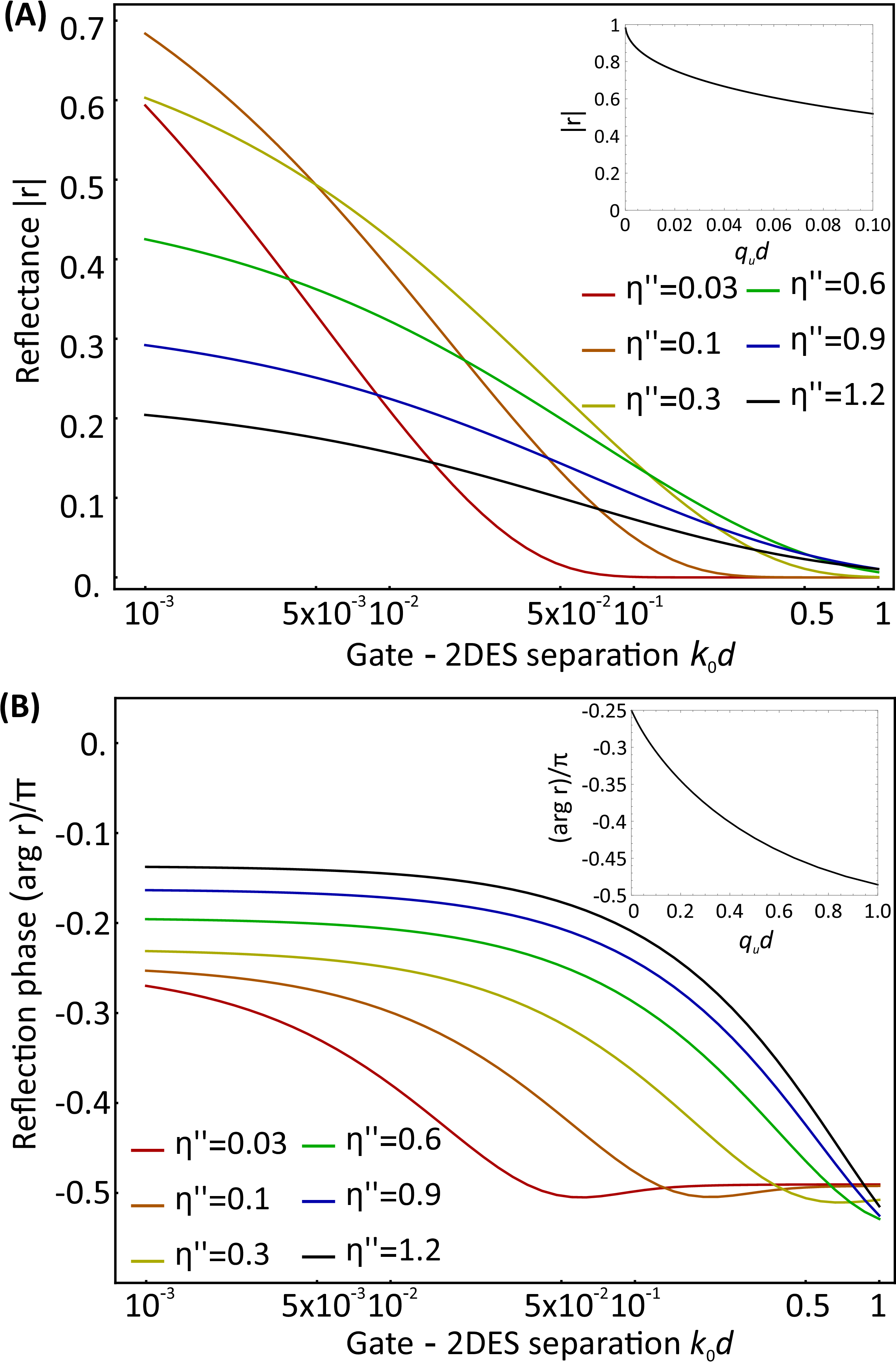}}
\caption{Reflection of the 2D plasmon at a single gate edge. 
 Magnitude of the reflection coefficient (a) and its phase (b), both plotted as functions of the normalized gate-2DES separation $k_0 d$, where $k_0$ is the free-space wave vector. Different colors correspond to different values of normalized conductivity, $\eta$, which is assumed purely imaginary ($\{\eta' = 10^{-5}\} \ll \eta''$). Insets show the amplitude and phase of reflection in the non-retarded limit, which depend now only on gate-2DES separation normalized by ungated plasmon wave vector $q_ud$} 
\label{Fig-reflectance}
\end{figure}

Equation (\ref{eq-reflectance-full}) for the plasmon reflectance at the boundary between gated and ungated regions is one of our central results. It fully accounts for evanescent waves excited near the gate boundary, as well as plasmon radiative losses, i.e. the emission of free-space electromagnetic waves upon scattering. The resulting dependences of absolute reflectance $|r|$ and the phase ${\rm arg}\,r$ are shown in Fig.~\ref{Fig-reflectance} at different values of gate-2DES separation (normalized as $k_0d$) and normalized conductivity $\eta$. 

Considerable simplification of (\ref{eq-reflectance-full}) is possible for weakly dissipative 2DES, $\eta''\gg \eta'$, which is the only case of interest for plasmonics. In this case, the functions $\varepsilon_{u/g}$ have finite imaginary part only for 'radiative' wave vectors $-k_0 < q < k_0$. Otherwise, these functions are real and change sign at $q=\pm q_{u/g}$. This knowledge is sufficient to extract the absolute value and phase for the split functions ($\alpha=\{u,g\}$ distinguishes between ungated and gated functions):
\begin{gather}
\varepsilon_{\alpha+}(q) = \sqrt{\varepsilon_\alpha(q)\frac{q_\alpha+q}{q_\alpha - q}}\varepsilon_{\alpha,\rm rad}(q)e^{i\phi_\alpha(q)/2},\\
\varepsilon_{\alpha,\rm rad}(q) = \exp\left\{ \frac{1}{2\pi} \int\limits_{-k_0}^{k_0} \arctan\left[\frac{\varepsilon_\alpha''(u)}{\varepsilon_\alpha'(u)}\right]\frac{du}{u-q} \right\},\\
\label{eq-phase-int}
\phi_\alpha(q) = -\frac{1}{\pi} \int\limits_{-\infty}^{+\infty}{\ln\left|\frac{\varepsilon_\alpha(u)}{1-u^2/q_\alpha^2}\right|\frac{du}{u-q}}.
\end{gather}

\corra{The reflection phase is now linked to the phases of split functions:
\begin{equation}
    {\rm arg}r = \pi + \pi/2 + \phi_u(q_u) - \phi_g(q_u),
\end{equation}
where $\pi$ is the normal phase shift expected for reflection from opaque objects, $\pi/2$ is the additional phase shift provided by the sharp metal edge of the gate, and evanescent fields excited at the ungated and gated sides, respectively, are responsible for the shifts $\phi_u(q_u)$ and $-\phi_g(q_u)$.
}

\corra{Most plasmonic experiments are performed at $\eta'' \ll 1$, where the confinement is strong and plasma wave vector $q_u\approx k_0/\eta''$ is far above that of light. The limit is known as non-retarded one. In Appendix B, we derive a systematic expansion of $r$ in powers of $\eta''$, and show that (1) radiative loss is small, $\varepsilon_{u,\rm rad} \approx 1-\eta''^2/4$, $\varepsilon_{g,\rm rad}\approx1-(3/8){{{\eta''}}^{2}}{{\left( {{k}_{0}}d \right)}^{2}}$ (2) the phase shift due to ungated evanescent modes is approximated as $\phi_u(q_u) \approx \pi/4 - \pi^{-1}\eta''^2 \ln \eta''$, where the $\pi/4$ term was known previously~\cite{Khavasi_pi4_phase, Moreno_phase} and the presence of 'radiative correction' was noted in numerical experiments~\cite{Retardation_loss} (3) the only dimensionless parameter governing the complex reflection $r$ becomes $q_u d \approx k_0d/\eta''$. The non-retarded reflectance simplifies to 
\begin{equation}
\label{eq-reflectance-simple}
{r}=\exp \left[ - i \frac{\pi}{4} - i \phi_g (q_u) \right]\frac{1-{{q}_{u}}/{{q}_{g}}}{1+{{q}_{u}}/{{q}_{g}}} ,
\end{equation}
its argument and phase are shown in the insets of Fig.~\ref{Fig-reflectance} as functions of $q_ud$. The phase $\phi_g (q_u)$ interpolates between zero for proximate gates ($\phi_g(q_u) \approx \pi^{-1} q_ud [4-\ln(2q_ud)]$ at $q_ud\ll1$) and $\pi/4$ for distant gates ($q_dd\gg1$). In the most interesting case of proximate gate, the reflection phase equals $-\pi/4$. It is different from $+\pi/4$ phase shift from terminated 2DES edge~\cite{Khavasi_pi4_phase, Moreno_phase}. The physical origin of difference is singular enhancement of local fields by the metal gate edge.} The results in Fig.~\ref{Fig-reflectance} b confirm the anticipated shift: all curves coalesce to $-\pi/4$ at small conductivity $\eta''\ll1$ and small gate-2DES distance $q_ud\ll 1 $.

Knowing the reflection coefficient at a single gate edge above the 2DES, we proceed to the analysis of the resonant modes in a plasmonic slot cavity formed by parallel edges separated by a distance $L$ [Fig. \ref{structure} a]. We adopt a quasi-optical approach, in which the cavity mode is represented as a combination of forward and backward plane waves. By requiring the coincidence of the complex amplitudes after a cavity round-trip, one obtains the dispersion law:
\begin{equation}
\label{eq-eigenmodes}
    r^2 \exp\{2 i q_u(\omega) L\} = 1,
\end{equation}
where $r$ is given by (\ref{eq-reflectance-full}) or its simplified version (\ref{eq-reflectance-simple}). The  slot plasmon spectrum obtained from Eqs.~(\ref{eq-eigenmodes}) and (\ref{eq-reflectance-simple}) reads as 
\begin{gather}
    \omega_n = \omega'_n-i\omega''_n,\\
    \label{eq-eigenmodes-freqs}
    \omega'_n = \pi\left( n + \frac{1}{4}\right)\frac{c\eta''}{L},\\
    \label{eq-eigenmodes-freqs2}
    \omega''_n =\frac{c \eta''}{L}\ln\left[ \frac{1+{{q}_{u}}/{{q}_{g}}}{1-{{q}_{u}}/{{q}_{g}}} \right]+\frac{\omega'_n\eta'}{\eta''}.
\end{gather}
\corra{The above spectrum is asymptotically valid in the highly-confined limit $\eta''\ll 1$ and for proximate gates $q_ud \ll 1$.} \corrb{The presence of substrate below 2DES with dielectric constant $\epsilon_s$, typical for plasmonic experiments, can be taken into account via replacement $\eta \rightarrow \eta/\overline{\epsilon}$, $\overline{\epsilon} = (\epsilon_s+1)/2$.} \corra{Frequency dispersion of conductivity, if present, is not restrictive: substituting the known dispersive model $\eta(\omega)$ into (\ref{eq-eigenmodes}) and solving  with respect to $\omega$ would yield the modified spectrum. Particularly, for Drude model one would get $\omega'_n = \sqrt{2\pi n_se^2q_n/(\overline{\varepsilon} m)}$ with quantized wave vectors $q_n = (n+1/4)\pi/L$, where $n_s$ is the sheet density of 2d electrons and $m$ is the effective mass. }

The first remarkable property of the slot plasmon modes is the unconventional quantization rule (\ref{eq-eigenmodes-freqs}), where the mode numbers $n$ gain a constant offset of $1/4$. It is a direct consequence of non-trivial phase shift of $-\pi/4$ upon reflection of the wave from the gate edge. As a result, the fundamental resonance in the plasmonic cavity is excited when $L=\lambda_u/8$, which differs significantly from the familiar expression for frequency in an optical Fabry-Perot cavity, where $L=\lambda_0/2$. 

The second remarkable property of the slot plasmon modes is their quasi-bound nature and decay into plasmons under the gates. Formally, this is manifested by a finite decay rate (\ref{eq-eigenmodes-freqs2}) even for a clean 2DES with $\eta'\rightarrow 0$. The decay constant becomes small as the gate-2DES separation approaches zero, $\omega''_n \sim \omega'_n (q_u d)^{1/2}$. As $d$ decreases, the decay rate approaches zero quite slowly, implying that observing such modes requires very small separations between the gate and 2DES.

We proceed to compare our analytical results (\ref{eq-eigenmodes-freqs},\ref{eq-eigenmodes-freqs2}) with electromagnetic simulations. This comparison is necessary due to the approximate nature of the Fabry-Perot-type approach (\ref{eq-eigenmodes}) for the eigenmodes. Although it fully accounts for evanescent field effects at a single boundary (encoded in ${\rm arg}\, r$), it neglects the interactions of evanescent fields between the two boundaries. This neglect is well justified for high-order modes $n \gg 1$, whereas its applicability for $n\sim 1$ may be questionable. The situation is analogous to Bohr-Sommerfeld quantization in quantum mechanics, which fortunately works well even for $n\sim 1$, and we hope for a similar result in our case.

In a numerical experiment, we illuminate the slot above the 2DES with a normally incident electromagnetic wave [Fig.~\ref{structure} a] and study the electromagnetic absorption cross-section $A$ having the dimension of length. Simulations are performed in CST Microwave studio package. For simplicity, we use a frequency-independent 2DES conductivity $\eta(\omega) = {\rm const}$, although real conductivity functions are generally dispersive.

\begin{figure}
\center{\includegraphics[width=\linewidth]{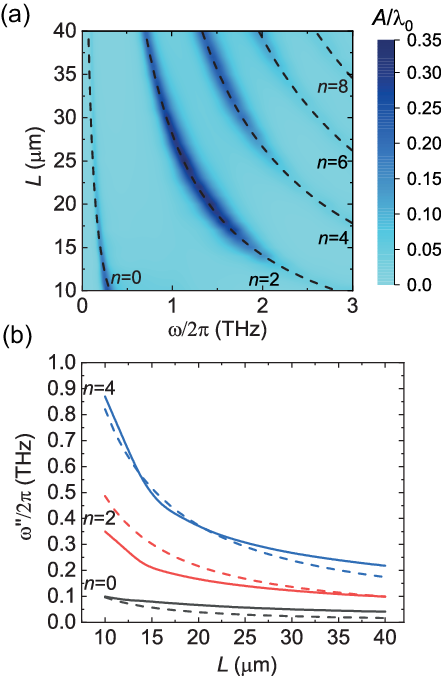}}
\caption{(a) The spectrum of absorption cross section $A$ normalized at free-space wavelength as function of slot width $L$. The dashed curves show the slot plasmon modes frequencies (Eq.~\ref{eq-eigenmodes-freqs}) 
as function of the slot width $W$ obtained analytically. (b) The linewidth of the plasmon modes $n=0, 2, 4$ calculated analytically using Eq.~\ref{eq-eigenmodes-freqs2} (dashed curves) and fitted from the simulations data (solid curves).}
\label{colormap_absorption}
\end{figure}

The results of absorption simulation are shown in Fig. \ref{colormap_absorption} (a) as a function of frequency and slot length $L$. The absorption peaks correspond to the excitation of cavity modes. Their frequency positions match very well the analytical theory [Eq. (\ref{eq-eigenmodes-freqs}), dashed black lines] if even $n$-values are used, $n=\{0,2,4...\}$. Odd-$n$ modes are dark, i.e. anti-symmetric with respect to electric field, and cannot be excited by a normally incident electromagnetic wave. Still, their excitation is possible at the inclined wave incidence (Appendix C).

The linewidth extracted from the simulations is shown in Fig.~\ref{colormap_absorption}(b) by the solid lines. The analytical theory is depicted in the same figure by the dashed lines. For both decay channels, i.e. leakage into gated modes and intrinsic 2DES dissipation, the linewidth grows with mode frequency. Such behavior can be interpreted recalling relatively short wavelength of high-order modes; as a result, they can easily 'sneak' under the gate.

A noteworthy property of slot plasmons observed in simulations is their large excitation cross-section. The fundamental dipole limit of absorption by resonant linear objects equals $2\lambda_0/\pi \approx 0.6\lambda_0$, where $\lambda_0$ is the {\it free space} wavelength ~\cite{Ruan2010}. The fundamental limit is achieved generally by  matching of the radiative and non-radiative losses~\cite{Mylnikov2022}. We observe excitation cross sections on the order of $(0.1...0.3)\lambda_0$, or $0.15...0.5$ of the dipole limit, without any special matching or optimization procedures. We attribute the efficient excitation of the slot modes to the two reasons. The former lies in singular enhancement of the electromagnetic field near the keen gate edges. The latter lies in the decay of slot modes into the propagating gated plasmons; such propagation involves the distant regions of 2DES into the absorption process.

In discussion, we suggest several consequences of our findings. First of all, the cavities formed by gate slots naturally resolve the problem of electromagnetic coupling to the deep subwavelength plasmonic structures~\cite{Mylnikov2022,Abajo_Limits}. The problem arose from small size of such objects and, hence, their weak dipole moment. Matching of such cavities required very clean 2D systems with conductivity $\eta'$ order of $Z_0^{-1}\lambda_0/\lambda_{\rm pl}$. In a slot cavity, no special matching conditions are required. 

Another applied consequence of our study is the formation of plasmonic 'hot spots' in gaps between gates and electrical contacts to low-dimensional nanostructures. These ubiquitously present gaps confine the incident radiation. Previous efforts on ultimate electromagnetic confinement and field enhancement in 2DES were concentrated on gated plasmon resonances~\cite{Iranzo2018,AlonsoCalafell2021}, while our findings show that field confinement with gap modes is no less promising.

Our results show that prevailing theories of plasmonic states in grating-gated 2DES, so-called plasmonic crystals, require substantial revision. Such theories have often relied on matching the amplitudes of plane plasma waves between gated and ungated regions\cite{Aizin_finite, Kachorovskii_gated_instability, Gorbenko_LateralPC, Miranda_2024}.  This approach neglects the phase shift of $-\pi/4$ acquired upon plasmon reflection from gated region, thereby systematically overestimating the eigenfrequencies of modes confined between gates. The omission long went unnoticed in spectroscopy, which predominantly probed either the weak-coupling regime~\cite{Allen1977, Theis_inversion, Khisameeva_crystal} or modes localized beneath individual gates~\cite{Bylinkin_tight_binding, Iranzo2018}. A class of slot modes was recently observed in a plasmonic crystal with a totally depleted regions beneath the gates~\cite{Sai2023}. The authors acknowledged, without offering an explanation, a strong disagreement of the observed resonant frequency with 'conventional' quantization rule. In an accompanying paper~\cite{Accompanying}, observation of slot plasmon modes is discovered for a plasmonic crystal with uniform 2DES, and a very good agreement with dispersion (\ref{eq-eigenmodes-freqs}) is found.

In summary, we have theoretically studied a new class of resonant cavities for 2D plasmons formed by slot in the metal gates. The unusual 'quantization rule' for the mode frequencies $L = \lambda/8 +n \times \lambda/2$, where $n=0, 1, 2 \ldots$ is a consequence of anomalous $-\pi/4$ phase shift of 2D plasmon reflecting from the gate edge. Large excitation cross section of slot plasmons approaching the dipole limit is facilitated by field enhancement at the gate edges.

This work was supported by the Russian Science Foundation (Grant No. 24-79-00094). The authors thank Denis Fateev for helpful discussions.

\begin{widetext}
\appendix
\section{Derivation of electromagnetic scattering equation}
Derivation of the governing equation for electromagnetic fields in the partly gated 2DES starts with superposition principle for vector potential ${\bf A}$. It is created by external sources, currents in 2DES, and currents in gate, respectively: 
\begin{equation}
	{\bf A}={{{\bf A}}_{ext}}+{\bf A}\left\{ {{\bf j}_{2d}} \right\}+{\bf A}\left\{ {{\bf j}_{g}} \right\}
\end{equation}
Explicit formulation is most convenient in the mixed representation, where the dependence of all fields on the vertical $z$ coordinate is retained, and the Fourier transform is performed with respect to the $x$-coordinate
\begin{equation}
	{\bf A}(q,z) = \int_{-\infty}^{+\infty}{{\bf A}(x,z)e^{-iqx}dx}.
\end{equation}
In this representation, the link between surface currents and vector potential is given by:
\begin{equation}
	{\bf A}\{{\bf j}_{i}(q)\}(q,z)=\frac{2\pi }{c\kappa \left( q \right)}{{\bf j}_{i}(q)}{{e}^{-\kappa \left( q \right)\left| z-{{z}_{i}} \right|}}
\end{equation} 	
where the index $i=\{2d,g\}$ distinguishes between currents in the 2DES and gate, ${{z}_{i}}$ is the location of current-carrying plane. For brevity, we introduce 
\begin{equation}G\left( q \right)=2\pi /c\kappa \left( q \right),
\end{equation}
the Fourier transform of the fundamental solution of the wave equation. With this aid, we write for electric potential at any point $z$ in space:
\begin{equation}
	{{\bf A}}\left( q,z \right)={{{\bf A}}_{ext}}\left(q, z \right)+{{{\bf j}}_{2d}}(q)G\left( q \right)e^{-\kappa \left( {\bf q} \right)|z-z_{2d}|} +{{{\bf j}}_{g}}(q)G\left( q \right){{e}^{-\kappa \left( {\bf q} \right)|z-z_g|}}.
\end{equation} 
We will further choose the coordinate system with $z_{2d}=0$, $z_g=d$.

We proceed to the closed-form equation for fields in the gate plane. From now on, we choose the TM polarization of all fields, i.e. the vector potential and electric field have only $x$ and $z$ components, ${\bf A} = \{A_x,0,A_z\}$, ${\bf E} = \{E_x,0,E_z\}$. The $x$-component of the electric field will be of principal interest for us, as the currents in 2DES and in the gate possess the $x$-component only, ${\bf j}_g=\{j_{g,x},0,0\}$, ${\bf j}_{2d}=\{j_{2d,x},0,0\}$. The electrodynamic quantities from now on will correspond to the $x$-components, $E_x\equiv E$, $j_{g,x}\equiv j_g$, $j_{2d,x}\equiv j_{2d}$.

We relate the 2d current density to the electric field with Ohm's law
\begin{equation}
	{{ j}_{2d}}(q)={\sigma }{ E}\left(q, z=0 \right).
\end{equation}
The electric field is expressed via vector potential
\begin{equation}
	\label{eq-EA}
	{E}(q,z)=\frac{i}{{{k}_{0}}}\left( k_{0}^{2}-{{q}^{2}} \right){ A}(q,z)=-\frac{i}{{{k}_{0}}}{{\kappa }^{2}}\left( q \right)A(q,z).
\end{equation}
Upon deriving (\ref{eq-EA}), we have used ${\bf E}=-{\boldsymbol{\nabla}}\varphi (q) +i{{k}_{0}}{\bf A}$, while $\varphi$ was obtained from the Lorentz gauge 
\begin{equation}
	\varphi =-\frac{i}{{k_{0}}} {\nabla {\bf A}} .
\end{equation}
Introducing the 2DES current density via vector potential, we find
\begin{equation}
	{{A}}\left(q, z=0 \right)={{{ A}}_{ext}}\left( q,z=0 \right)-\frac{i{{\sigma }_{2d}}}{{{k}_{0}}}{{\kappa }^{2}}\left( q \right)G\left( q \right){{ A}}\left(q,z= 0 \right)+j_g(q)G\left( q \right){{e}^{-\kappa \left( {\bf q} \right)d}},
\end{equation} 
from which the vector-potential and current in the 2DES plane are obtained:
\begin{gather}
	\label{eq-link-g-2d}
	{{ A}}\left( q,z=0 \right)\varepsilon \left( q \right)={{ A}_{ext}}\left(q,z = 0 \right)+{j}_{g}(q)G\left( q \right){{e}^{-\kappa \left( {\bf q} \right)d}}, \\
	{j}_{2d}(q)=-\frac{i{{\sigma }_{2d}}}{{{k}_{0}}}\frac{{{\kappa }^{2}}\left( q \right)}{\varepsilon \left( q \right)}\left[ {{ A}_{ext}}\left(q,z= 0 \right)+{{ j}_{g}(q)}G\left( q \right){{e}^{-\kappa \left( {\bf q} \right)d}} \right].   
\end{gather}
The screening function is
\begin{equation}
	\varepsilon \left( q \right)=1+\frac{i{{\sigma }_{2d}}}{{{k}_{0}}}{{\kappa }^{2}}\left( q \right)G\left( q \right)=1+\frac{2\pi i{{\sigma }_{2d}}}{c}\frac{\kappa \left( q \right)}{{{k}_{0}}}.
\end{equation} 
Now we rewrite the fundamental solution of wave equation in the gate plane $z=d$:
\begin{equation}
	{{A}}\left(q,z= d \right)={{ A}_{ext}}\left(q,z= d \right)+{{ j}_{2d}}G\left( q \right){{e}^{-\kappa \left( {\bf q} \right)d}}+{j}_{g}(q)G\left( q \right)
\end{equation} 
and introduce the 2d current density hereby
\begin{multline}
	{{ A(q,z=d)}={ A}_{ext}}\left( q,z=d \right)-\\
	\frac{i{{\sigma }_{2d}}}{{{k}_{0}}}\frac{{{\kappa }^{2}}\left( q \right)}{\varepsilon \left( q \right)}\left[ {{A}_{ext}}\left(q,z= 0 \right)+{{ j}_{g}(q)}G\left( q \right){{e}^{-\kappa \left( {\bf q} \right)d}} \right]G\left( q \right){{e}^{-\kappa \left( {\bf q} \right)d}}+{{j}_{g}(q)}G\left( q \right)
\end{multline} 

The last remaining step is to relate the vector-potentials at two different $z$ coordinates. This is achieved with
\begin{equation}
	{{A}_{ext}}\left(q,z=0 \right)={{ A}_{ext}}\left(q,z=d \right){{e}^{i{{k}_{z}}\left( q \right)d}}.
\end{equation} 	
Afterwards, the scattering equation becomes 
\begin{equation}
	{A}\left( q,z=d \right)={{A}_{ext}}\left( q,z=d \right)\frac{{{\varepsilon }_{g}}\left( q \right)}{\varepsilon \left( q \right)}+{{j}_{g}}\left( q \right)G\left( q \right)\frac{{{\varepsilon }_{g}}\left( q \right)}{\varepsilon \left( q \right)}
\end{equation} 
An alternative representation is derived by expressing all vector potentials via the respective electric fields, Eq.~\ref{eq-EA}:

\begin{equation}
	\label{eq-e-equation}
	{E}\left( q,z=d \right)={{ E}_{ext}}\left( q,z=d \right)\frac{{{\varepsilon }_{g}}\left( q \right)}{\varepsilon \left( q \right)}-\kappa \left( q \right){{j}_{g}}\left( q \right)\frac{2\pi i}{c{{k}_{0}}}\frac{{{\varepsilon }_{g}}\left( q \right)}{\varepsilon \left( q \right)}
\end{equation}

This is the electromagnetic scattering equation we shall further solve for  plasmon reflection upon scattering at the gate edge.

The solution of the plasmon scattering problem is achieved by setting the external field to zero, $E_{\rm ext}\equiv 0$ in Eq.~\ref{eq-e-equation}. Instead, the total field is presented as the sum of incident plasma wave and the scattered field:
\begin{equation}
	\label{eq-total}
	E(q,z) = E_{\rm inc}(q,z) + E_{\rm scat} (q,z).
\end{equation}
The incident field in the real-space representation is a running exponent with wave vector $q_u$, bounded to $x<0$:
\begin{equation}
	E_{\rm inc}(x,z) = E_0(z)e^{iq_u x}\theta(-x).
\end{equation}
After the Fourier transform, it becomes 
\begin{equation}
	\label{eq-incident}
	E_{\rm inc}(q,z) = \frac{iE_0(z)}{q-q_u}.
\end{equation}
Denoting $E_0(z) \equiv E_0$ and introducing the representations (\ref{eq-incident}) and (\ref{eq-total}) into (\ref{eq-e-equation}), we arrive at the scattering equation (1) of the main text:
\begin{equation}
	\left[ \frac{i{{E}_{0}}}{q-{{q}_{u}}}+{{E}_{\rm scat}}\left( q \right) \right]\frac{\varepsilon_u \left( q \right) }{{{\varepsilon }_{g}}\left( q \right)}\frac{k_0}{\kappa(q)} + \frac{i}{2} j_g(q) Z_0=0.
\end{equation}


\section{Analytical approximations to the reflection coefficient}
\subsection{General simplifications}
Having solved the Wiener-Hopf problem, we obtained the expression for the scattered field:
\begin{gather}
	{{E}_{\rm scat}}\left( q \right) = -\frac{i{{E}_0}}{q-{q_u}}\left[ 1-\frac{{M}_{+}\left( q_u \right)}{{{M}_{+}}\left( q \right)} \right],\\
	M\left( q \right)=\frac{\varepsilon_u \left( q \right)}{{{\varepsilon }_{g}}\left( q \right)\kappa \left( q \right)}.
\end{gather}
To obtain the real-space field, we perform the inverse Fourier transform:
\begin{equation}
	{{E}_{\rm scat}}\left( x \right)=(2\pi)^{-1}\int\limits_{-\infty}^{+\infty}{{{E}_{\rm scat}}\left( q \right) e^{iqx}dq}.
\end{equation}
We close the integration loop in the {\it lower} half-plane of complex $q$-variable, where the exponent $e^{iqx}$ decays rapidly at $x<0$. The integration loop bypasses the branch cut of $\kappa(q)=\sqrt{q^2-k_0^2}$, which is a straight line starting at $k_0=-\omega/c-i\delta$ and running to $-i\infty$. Inside of this loop, there is a single pole of the integrand ${{E}_{\rm scat}}\left( q \right)$. It is located at $\varepsilon_u(q)=0$, or, equivalently, at $q=-q_u$. The contribution of this pole to the total field is exactly the field of the reflected plasma wave. Using the residue theorem, we find:
\begin{equation}
	{{E}_{r}}\left( x \right)={{e}^{-i{{q}_{u}}x}}\frac{{{E}_{0}}}{(-q_u)-{{q}_{u}}}\underset{q=-q_u}{ \rm{Res}}\,\frac{{{M}_{+}}\left( {{q}_u} \right)}{{{M}_{+}}\left( q \right)}
\end{equation}

Evaluation of residue is achieved by expanding $\varepsilon_u(q)$ near $q=-q_u$. This leads us to:
\begin{equation}
	{{E}_{r}}\left( x \right)={{e}^{-i{{q}_u}x}}\frac{{{E}_0}}{-2q_u}\frac{{{M}_{+}}\left( {{q}_u} \right){{\varepsilon }_{g}}\left( -q_u \right)\kappa \left( - q_u \right){{M}_{-}}\left( -q_u \right)}{{{\left. \partial \varepsilon /\partial q \right|}_{q={{q}_{\rm{refl}}}}}}
\end{equation}
Expression for the reflection coefficient (10) from the main text follows after two simplifications. First, ${{f}_{+}}\left( q \right)={{f}_{-}}\left( -q \right)$, which holds for any 'plus' and 'minus' factorized functions as a consequence of Cauchy factorization. Second, the value $\kappa_+(q_u)$ is obtained using the rules of analytical continuation for square root function. The physical choice is $\kappa(q=0)=-ik_0$, thus $\kappa_+(q=0)=\sqrt{-ik_0}$. Continuing the function to $q=-q_u$, we find
\begin{equation}
	{{\kappa }_{+}}\left( {{q}_{\rm{pl}}} \right)=\sqrt{-i{{k}_{0}}}\sqrt{\left| \frac{{{q}_{pl}}+{{k}_{0}}}{{{k}_{0}}} \right|}{{e}^{i\Delta \arg /2}},
\end{equation} 	
where $\Delta \arg=\pi$ is the change of the argument.

The above preliminaries allow us to find:
\begin{equation}
	{r}=-\frac{i}{2{q_u}}{{\left[ \frac{{{\varepsilon }_{u+}}\left( {{q}_{u}} \right)}{{{\varepsilon }_{g+}}\left( {{q}_{u}} \right)} \right]}^{2}}\sqrt{\frac{{{q}_{u}}/{{k}_{0}}-1}{{{q}_{u}}/{{k}_{0}}+1}}\frac{e^{ -2 \kappa(q_u)d }}{{{\left. \partial \varepsilon_u /\partial q \right|}_{q=-{q_u}}}}.
\end{equation}

Further analysis is simplified in the weakly dissipative limit, $\eta'\ll\eta''$. In that case, the screening functions $\varepsilon_{\alpha}(q)$, $\alpha=\{u,g\}$ have finite imaginary part only for $-k_0 < q < k_0$. They also change sign at $q=\pm q_{\alpha}$. We introduce the auxiliary screening functions $\varepsilon_{\alpha}^{aux}(q)$ that have simpler analytic structure and the same zeros as $\varepsilon_{\alpha}(q)$:
\begin{equation}
	\varepsilon^{aux}_{\alpha}(q) = 1 - \frac{q^2}{q_{\alpha}^2}.
\end{equation}
Auxiliary functions are factorized immediately as they are polynomials of $q$. Factorization of the full screening functions can be now presented as:
\begin{equation}
	\varepsilon_{\alpha\pm}(q) = \left(1\pm\frac{q}{q_\alpha}\right) \left[ \frac{\varepsilon_\alpha(q)}{\varepsilon^{aux}_\alpha(q)}\right]_\pm.
\end{equation}
We now apply the Cauchy factorization procedure to the function $\varepsilon_\alpha(q)/\varepsilon^{aux}_\alpha(q)$:
\begin{equation}
	\left[ \frac{\varepsilon_\alpha(q)}{\varepsilon^{aux}_\alpha(q)}\right]_\pm  =\exp \left\{ \pm \frac{1}{2\pi i}\int\limits_{-\infty }^{+\infty }{\frac{\ln \left[ \frac{\varepsilon_\alpha(u)}{\varepsilon^{aux}_\alpha(u)}\right]du}{u-\left( q\pm i\delta  \right)}} \right\}.
\end{equation} 
We apply the Sokhotski theorem to the integral, keeping in mind that we shall be further interested in values of $q$ close to the real axis:
\begin{equation}
	\left[ \frac{\varepsilon_\alpha(q)}{\varepsilon^{aux}_\alpha(q)}\right]_\pm  =\sqrt{\frac{\varepsilon_\alpha(q)}{\varepsilon^{aux}_\alpha(q)}}\exp \left\{ \pm \frac{1}{2\pi i}{\rm v.p.}\int\limits_{-\infty }^{+\infty }{\frac{\ln \left[ \frac{\varepsilon_\alpha(u)}{\varepsilon^{aux}_\alpha(u)}\right]du}{u-q}} \right\}.
\end{equation} 
This function $\varepsilon_\alpha(q)/\varepsilon^{aux}_\alpha(q)$ under the logarithm has no zeros, thus the logarithm is real-valued at all values of $q$, except for the 'radiative' values $-k_0<q<k_0$. The 'radiative' values of $q$ contribute to the modulus of the factorized function, while the 'non-radiative' values do not. Splitting the integral 
\begin{equation}
	\int\limits_{-\infty }^{+\infty }...=\int\limits_{-\infty }^{-k_0 }... + \int\limits_{-k_0 }^{+k_0 }...+\int\limits_{+k_0 }^{+\infty }...,
\end{equation}
and expressing explicitly the imaginary part of logarithm:
\begin{equation}
	{\rm Im}\ln(\varepsilon'_\alpha+i\varepsilon''_\alpha) = {\rm arctan}\frac{\varepsilon_\alpha''}{\varepsilon_\alpha'}
\end{equation}
we arrive at
\begin{equation}
	\varepsilon_{\alpha\pm}(q) = \sqrt{\varepsilon_\alpha(q)\frac{q_\alpha\pm q}{q_\alpha \mp q}}\exp\left\{ \frac{1}{2\pi} \int\limits_{-k_0}^{k_0} \arctan\left[\frac{\varepsilon_\alpha''(u)}{\varepsilon_\alpha'(u)}\right]\frac{du}{u-q} \right\} \exp\left\{\mp\frac{i}{2 \pi} \int\limits_{-\infty}^{+\infty}{\ln\left|\frac{\varepsilon_\alpha(u)}{1-u^2/q_\alpha^2}\right|\frac{du}{u-q}}\right\},
\end{equation}
which is equivalent to Eqs.~(11)-(13) of the main text. 

\subsection{Reflection phases: analytical considerations}

We proceed now to the detailed evaluation of the phase of split functions:
\begin{equation}
	\label{eq-phase-gen}
	\phi_\alpha(q) = -\frac{1}{\pi} \int\limits_{-\infty}^{+\infty}{\ln\left|\frac{\varepsilon_\alpha(u)}{1-u^2/q_\alpha^2}\right|\frac{du}{u-q}}.
\end{equation}

\subsubsection{Phase due to the ungated evanescent fields}
We start from the ungated function $\phi_u(q_u)$ that depends on a sole parameter $\eta''$. Our aim is to obtain the systematic expansion of $\phi_u(q_u)$ in powers of $\eta ''\ll 1$. The ungated wave vector $q_u/k_0 =(1 +1/\eta''^2)^{1/2}$. Splitting the phase integral into radiative and non-radiative parts, we find
\begin{gather}
	{{\phi }_{u}}\left( {{q}_{u}} \right)={{I}_{1}}+{{I}_{2}} \\ 
	{{I}_{1}}\left( {\eta ''} \right)=\frac{1}{2\pi }\int\limits_{-1}^{1}{\ln \left( 1+\eta ''^2\left( 1-{{u}^{2}} \right) \right)\frac{du}{u-{{q}_{u}}/{{k}_{0}}}} \\ 
	{{I}_{2}}\left( {\eta ''} \right)=\frac{2}{\pi }\int\limits_{1}^{+\infty }{\ln \left( 1+\eta ''\left( 1-{{u}^{2}} \right)^{1/2} \right)\frac{udu}{{{u}^{2}}-{{\left( {{q}_{u}}/{{k}_{0}} \right)}^{2}}}} \\ 
\end{gather}

In the radiative part and in the non-retarded limit $\eta''\ll 1$, we can approximate 
\begin{equation}
	{{I}_{1}}\left( {\eta ''} \right)\approx \frac{1}{2\pi }\int\limits_{-1}^{1}{\eta ''^2\left( 1-{{u}^{2}} \right)\frac{du}{u-{{q}_{u}}/{{k}_{0}}}}\approx -\frac{2{{{\eta ''}}^{3}}}{3\pi}
\end{equation}

We shall see that the main contribution to the phase in the non-retarded limit comes from $I_2$. Making the change of variables $t = \eta'' (1-u^2)^{1/2}$, we rewrite
\begin{equation}
	{{I}_{2}}\left( {\eta ''} \right)=\frac{2}{\pi }\sqrt{1+{{{\eta ''}}^{2}}}\int\limits_{0}^{+\infty }{\frac{\ln \left( 1+t \right)}{\sqrt{{{t}^{2}}+{{{\eta ''}}^{2}}}}\frac{tdt}{{{t}^{2}}-1}}  
\end{equation}
First of all, we evaluate the leading term in the absence of retardation:
\begin{equation}
	{{I}_{2}}\left( 0 \right)=\frac{2}{\pi }\int\limits_{0}^{+\infty }{\ln \left( 1+t \right)\frac{dt}{{{t}^{2}}-1}}=\frac{\pi }{4}    
\end{equation}

The 'radiative corrections' are obtained by examining the difference ${{I}_{2}}\left( {\eta ''} \right)-{{I}_{2}}\left( 0 \right)$. We shall see that these corrections will be order of $\eta''^2 \ln(1/\eta'')$. With the leading-log accuracy, we can neglect the overall prefactor  $\sqrt{1+\eta''^2}$ and write
\begin{equation}
	{{I}_{2}}\left( \eta'' \right)-{{I}_{2}}\left( 0 \right)=\frac{2}{\pi }\int\limits_{0}^{+\infty }{\ln \left( 1+t \right)\left[ \frac{1}{\sqrt{{{t}^{2}}+{{\eta''}^{2}}}}-\frac{1}{t} \right]\frac{tdt}{{{t}^{2}}-1}}
\end{equation}

Two characteristic integration domains can now be singled out, $t\in[0, \eta'']$ and $t\in[\eta'',+\infty]$
\begin{multline}
	{{I}_{2}}\left( \eta'' \right)-{{I}_{2}}\left( 0 \right)=\\
	\frac{2}{\pi }\int\limits_{0}^{\eta''}{\ln \left( 1+t \right)\left[ \frac{1}{\sqrt{{{t}^{2}}+{{\eta''}^{2}}}}-\frac{1}{t} \right]\frac{tdt}{{{t}^{2}}-1}}+\frac{2}{\pi }\int\limits_{\eta''}^{+\infty }{\ln \left( 1+t \right)\left[ \frac{1}{\sqrt{{{t}^{2}}+{{\eta''}^{2}}}}-\frac{1}{t} \right]\frac{tdt}{{{t}^{2}}-1}}=\\
	A+B 
\end{multline}

In the domain of small $t<\eta''$, we expand the logarithm and approximate
\begin{equation}
	A\approx -\frac{2}{\pi }\int\limits_{0}^{\eta''}{\left[ \frac{t}{\sqrt{{{t}^{2}}+{{\eta''}^{2}}}}-1 \right]tdt}=-\frac{2{{\eta''}^{2}}}{\pi }\int\limits_{0}^{1}{\left[ \frac{x}{\sqrt{{{x}^{2}}+1}}-1 \right]xdx}.
\end{equation}
The $A$-term is order of $\eta''^2$ and does not yield the leading-log result. 

In the domain of large $t>\eta''$, we develop the series expansion of the square roots
\begin{equation}
	B=\frac{2}{\pi }\int\limits_{\eta''}^{+\infty }{\ln \left( 1+t \right)\left[ \frac{1}{\sqrt{{{t}^{2}}+{{\eta''}^{2}}}}-\frac{1}{t} \right]\frac{tdt}{{{t}^{2}}-1}}\approx \frac{2}{\pi }\int\limits_{\eta''}^{+\infty }{\ln \left( 1+t \right)\left[ -\frac{{{\eta''}^{2}}}{2{{t}^{3}}}+\frac{3{{\eta''}^{4}}}{8{{t}^{5}}}-\frac{5{{\eta''}^{6}}}{16{{t}^{7}}}+... \right]\frac{tdt}{{{t}^{2}}-1}}.
\end{equation}
The first term of expansion in the square brackets diverges logarithmically at the lower limit and yields the necessary leading result
\begin{equation}
	B\approx \frac{2}{\pi }\int\limits_{\eta''}^{{{t}_{\max }}}{\frac{{{\eta''}^{2}}}{2{{t}^{3}}}{{t}^{2}}dt}\approx \frac{1}{\pi }{\eta''^{2}}\ln \frac{1}{\eta''}.
\end{equation}
The precise value of $t_{\max}$ cannot be established with that method and is beyond the leading log approximation, we can set it to unity.

Collecting all relevant expansion terms, we find the phase of ungated function with leading-log radiative corrections
\begin{equation}
	\label{eq-ungated-phase-lead-log}
	{{\phi }_{u}}\left( {{q}_{u}} \right)=\frac{\pi }{4}+\frac{1}{\pi }{\eta''^{2}}\ln \frac{1}{\eta''}.
\end{equation}

\subsubsection{Phase due to the gated evanescent fields}
The phase of gated split function will be handled only in the non-retarded limit, where it can be presented as
\begin{equation}
	\varepsilon_g = 1- |\xi|(1-e^{-2|\xi|D}),
\end{equation}
where we have introduced the dimensionless wave vector $\xi=q/q_u$ and dimensionless gate-channel separation $D=q_u d$. The phase $\phi_g (q_u)$ which affects the reflection coefficient of plasmon, depends on a sole dimensionless parameter $D$ and is given by the following integral
\begin{equation}
	\label{eq-g-phase}
	\phi_g (q_u) = -\frac{1}{\pi} \int\limits_{-\infty}^{+\infty}{\frac{d\xi}{\xi-1}\ln\frac{1-|\xi|(1-e^{-2D|\xi|})}{1-\xi^2/\xi^2_g(D)}}.
\end{equation}
Here $\xi_g(D)$ is the dimensionless zero of the gated dielectric function, i.e. the solution of $1- |\xi_g|(1-e^{-2|\xi_g|D})=0$. In the limit of proximate gate $\xi_g(D) \approx1/\sqrt{2D} \gg 1$. It is easy to show that $\phi_g (q_u)\rightarrow 0$ as $D\rightarrow 0$. Indeed, the gated dielectric function is well approximated by polynomial $\varepsilon_g^{aux}(q)$ in a very broad range of wave vectors in that limit, and the logarithm in (\ref{eq-phase-gen}) is close to zero.

The leading corrections to $\phi_g (q_u)$ come from large $\xi$, such that $|\xi|>\{\xi_g(D)=1/\sqrt{2D}\}$. Only in that limit the full dielectric function $\varepsilon_g(q)$ deviates markedly from its polynomial approximation $\varepsilon_g^{aux}(q)$. The integral over large-$\xi$ domain can be estimated as
\begin{equation}
	\label{eq-gated-phase-lead-log}
	\phi_g(D) \approx \frac{D}{\pi}  (4-\ln2D).
\end{equation}

Comparison of exact results for phases $\phi_u$ and $\phi_g$ with their leading-log analytical approximations (\ref{eq-ungated-phase-lead-log}) and (\ref{eq-gated-phase-lead-log}) is shown in Fig.~\ref{Fig_S_phases}. As suggested, approximate expressions describe the behavior of phases accurately at small $\eta''$ and small $q_u d$, respectively.

\begin{figure}
	\center{\includegraphics[width=0.9\linewidth]{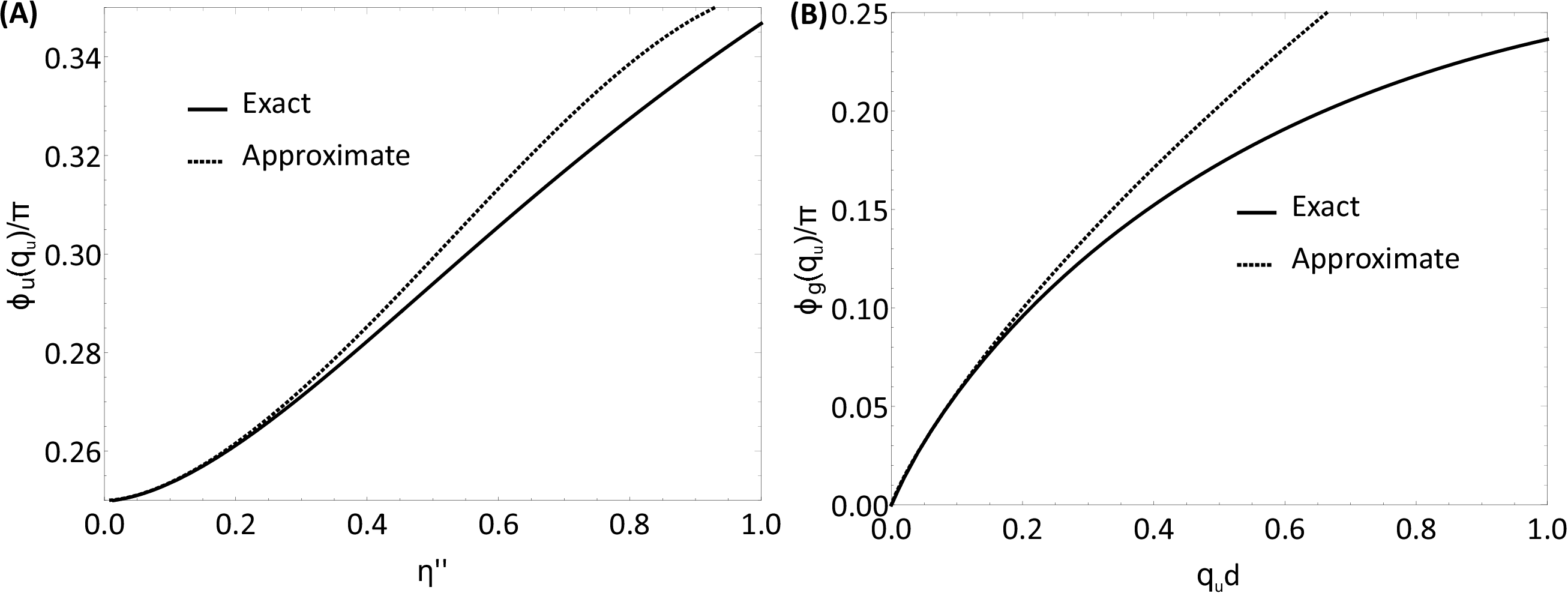}}
	\caption{Comparison of phases of split functions calculated with exact and approximate expressions (A) phase of the ungated function $\phi_u(q_u)$ vs imaginary part of conductivity $\eta''$ (B) phase of the gated function $\phi_g(q_u)$ vs normalized gate-2DES separation $q_ud$. Solid lines show the result for exact split functions, dashed lines represent analytical approximations using eqs. (\ref{eq-ungated-phase-lead-log}) and (\ref{eq-gated-phase-lead-log})}
	\label{Fig_S_phases}
\end{figure}

\subsection{Radiation corrections to absolute reflection: analytical considerations}
We proceed to evaluation of 'radiation corrections' to the absolute values of split functions, i.e. the functions
\begin{equation}
	\label{eq-rad-correction}
	\varepsilon_{\alpha,\rm rad}(q) = \exp\left\{ \frac{1}{2\pi} \int\limits_{-k_0}^{k_0} \arctan\left[\frac{\varepsilon_\alpha''(u)}{\varepsilon_\alpha'(u)}\right]\frac{du}{u-q} \right\}
\end{equation}

The wave vectors in (\ref{eq-rad-correction}) are bounded between the light cone $-k_0 <u <k_0$. At very small 2DES conductivity, $\eta''\ll1$, the functions $\varepsilon_{\alpha,\rm rad}(q)$ tend to unity. This justifies the notion of 'radiative corrections' for the above functions. Physically, they are responsible for the emission of free-space electromagnetic waves upon plasmon scattering at the edge. Naturally, they have nothing to deal with radiative corrections to energy levels and scattering cross-sections in quantum field theory.

\subsubsection{Radiation corrections to the ungated split function}
We start with radiation correction to the ungated split function $\varepsilon_{u,\rm rad}(q_u)$. It is given by the integral
\begin{equation}
	{{\varepsilon }_{u,\text{rad}}}({{q}_{u}})=\exp \left\{ \frac{1}{2\pi }\int\limits_{-1}^{1}{\arctan }\left[ \eta''\sqrt{1-{{u}^{2}}} \right]\frac{du}{u-{{q}_{u}}/{{k}_{0}}} \right\}.
\end{equation}

We denote the non-trivial integral in the exponent as $I\left( {\eta''} \right)$:
\begin{equation}
	I\left( {\eta''} \right)=\int\limits_{-1}^{1}{\arctan }\left[ \eta''\sqrt{1-{{u}^{2}}} \right]\frac{du}{u-{{q}_{u}}/{{k}_{0}}}=-{{\eta''}^{2}}\sqrt{1+\frac{1}{{{{\eta''}}^{2}}}}\int\limits_{-1}^{1}{du\frac{\arctan \left[ \eta''\sqrt{1-{{u}^{2}}} \right]}{{{{\eta''}}^{2}}\left( 1-{{u}^{2}} \right)+1}}
\end{equation} 

For actual evaluation, we introduce an auxiliary function $J\left( \mu ,\nu  \right)$
\begin{equation}
	J\left( \mu ,\nu  \right)=\int\limits_{-1}^{1}{du\frac{\arctan \left[ \mu \sqrt{1-{{u}^{2}}} \right]}{{{\nu }^{2}}\left( 1-{{u}^{2}} \right)+1}}
\end{equation}
which is linked to the original integral as
\begin{equation}
	I\left( {\eta''} \right)={{\eta''}^{2}}\sqrt{1+\frac{1}{{{{\eta''}}^{2}}}}J\left( \eta'',\eta'' \right).
\end{equation}

Differentiation of auxiliary function with respect to $\mu$ yields
\begin{equation}
	\frac{\partial J}{\partial \mu }=\int\limits_{-1}^{1}{du\frac{\sqrt{1-{{u}^{2}}}}{\left[ 1-{{\nu }^{2}}\left( 1-{{u}^{2}} \right) \right]\left[ 1-{{\mu }^{2}}\left( 1-{{u}^{2}} \right) \right]}=\frac{\pi }{{{\nu }^{2}}-{{\mu }^{2}}}\left[ \frac{1}{\sqrt{1+{{\nu }^{2}}}}-\frac{1}{\sqrt{1+{{\mu }^{2}}}} \right]}.
\end{equation}

Reverse integration is performed with relative ease:
\begin{equation}
	J\left( \nu ,\nu  \right)=\int\limits_{0}^{\nu }{d\mu \frac{\pi }{{{\nu }^{2}}-{{\mu }^{2}}}\left[ \frac{1}{\sqrt{1+{{\nu }^{2}}}}-\frac{1}{\sqrt{1+{{\mu }^{2}}}} \right]=\frac{\pi }{2}\frac{\ln \left( 1+{{\nu }^{2}} \right)}{\nu \sqrt{1+{{\nu }^{2}}}}}.
\end{equation}

Eventually, we get the integral under the exponential in the expression for radiative correction:
\begin{equation}
	I\left( {\eta''} \right)=-\frac{\pi }{2}\ln \left( 1+{{{\eta''}}^{2}} \right),
\end{equation}
the radiative correction itself
\begin{equation}
	{{\varepsilon }_{u,\text{rad}}}({{q}_{u}})={{\left[ 1+{{{\eta''}}^{2}} \right]}^{-1/4}}
\end{equation}
and the absolute value of the split function $\left| {{\varepsilon }_{u+}}({{q}_{u}}) \right|$ appearing in the expression for plasmon reflectance
\begin{equation}
	\label{eq-abs-ungated}
	\left| {{\varepsilon }_{u+}}({{q}_{u}}) \right|=\underset{q\to {{q}_{u}}}{\mathop{\lim }}\,\sqrt{{{\varepsilon }_{u}}(q)\frac{{{q}_{u}}+q}{{{q}_{u}}-q}}{{\varepsilon }_{u,\text{rad}}}({{q}_{u}})=\sqrt{2}{{\left( 1+{{{\eta''}}^{2}} \right)}^{1/4}}
\end{equation}

\subsubsection{Radiation corrections to the gated split function}
Radiation corrections to the gated split functions can be estimated analytically only in the limit of small gate-channel separation $k_0d \ll 1$. The limit is of main practical interest in 2d plasmonics.

Evaluation of radiation correction requires the extraction of real and imaginary parts of gated dielectric function
\begin{equation}
	{{\varepsilon }_{g}}(q)=1+i\eta''\frac{\sqrt{k_{0}^{2}-{{q}^{2}}}}{{{k}_{0}}}\left( 1-\exp \left[ 2i\sqrt{k_{0}^{2}-{{q}^{2}}}d \right] \right)
\end{equation}
in the radiative domain $-k_0<q<k_0$. Again, we limit ourselves to the weakly dissipative 2DES conductivity $\eta' \ll \eta''$. Expanding the phase exponent, we see that the imaginary part of ${{\varepsilon }_{g}}(q)$ appears only at the second order in $k_0d$:
\begin{equation}
	{{\varepsilon }_{g}}(q)\approx 1+i\eta''\frac{\sqrt{k_{0}^{2}-{{q}^{2}}}}{{{k}_{0}}}\left( -2i\sqrt{k_{0}^{2}-{{q}^{2}}}d-\frac{1}{2}{{\left( 2i\sqrt{k_{0}^{2}-{{q}^{2}}}d \right)}^{2}} \right)
\end{equation}

The necessary ratio of imaginary and real parts is estimated as:
\begin{equation}
	\frac{{\varepsilon''_{g}}(q)}{{\varepsilon '_{g}}(q)}\approx 2\eta''{{\left( 1-\frac{{{q}^{2}}}{k_{0}^{2}} \right)}^{3/2}}{{\left( {{k}_{0}}d \right)}^{2}}
\end{equation}

Substituting the ratio into the expression for radiative correction and performing explicit integration, we get
\begin{multline}
	{{\varepsilon }_{g,\text{rad}}}({{q}_{u}})=\exp \left\{ \frac{1}{2\pi }\int\limits_{-1}^{1}{2\eta''{{\left( 1-{{u}^{2}} \right)}^{3/2}}{{\left( {{k}_{0}}d \right)}^{2}}\frac{du}{u-{{q}_{u}}/{{k}_{0}}}} \right\}=\\
	\exp \left\{ \eta''{{\left( {{k}_{0}}d \right)}^{2}}{{\left. \left[ \frac{3Q}{2}-{{Q}^{3}}+{{\left( {{Q}^{2}}-1 \right)}^{3/2}} \right] \right|}_{Q={{q}_{u}}/{{k}_{0}}}} \right\}.
\end{multline}
Upon performing the integration, we have used the fact that $q_u/k_0>1$.

We can explicitly observe that radiation correction to the gated dielectric function contains a small parameter $(k_0d)^2 \ll 1$. In the limit of strong confinement $\eta''\ll 1$, an extra small parameter appears, which is easily seen by expansion of exponential at large values of $q_u/k_0 \ll 1$
\begin{equation}
	{{\varepsilon }_{g,\text{rad}}}({{q}_{u}})\approx \exp \left\{ -\frac{3}{8}\frac{\eta''{{\left( {{k}_{0}}d \right)}^{2}}}{{{q}_{u}}/{{k}_{0}}} \right\}\approx \exp \left\{ -\frac{3}{8}{{{\eta''}}^{2}}{{\left( {{k}_{0}}d \right)}^{2}} \right\}.
\end{equation}

The above estimate allows us to conclude that radiation corrections to the gated dielectric function are negligible at $k_0d\ll 1$ and especially at $\eta '' \ll 1$. The complete absolute value of the gated split function equals

\begin{multline}
	\label{eq-abs-gated}
	\left| {{\varepsilon }_{g+}}({{q}_{u}}) \right|=\sqrt{{{\varepsilon }_{g}}(q_u)\frac{{{q}_{g}}+q_u}{{{q}_{g}}-q_u}}{{\varepsilon }_{g,\text{rad}}}({{q}_{u}})=\\
	\sqrt{\frac{q_g+q_u}{q_g-q_u}}e^{ -\kappa(q_u)d }\exp \left\{ \eta''{{\left( {{k}_{0}}d \right)}^{2}}{{\left. \left[ \frac{3Q}{2}-{{Q}^{3}}+{{\left( {{Q}^{2}}-1 \right)}^{3/2}} \right] \right|}_{Q={{q}_{u}}/{{k}_{0}}}} \right\}
\end{multline}

Both full numerical and analytical [eqs. (\ref{eq-abs-ungated}) and (\ref{eq-abs-gated})] results for absolute values of split functions are summarized in Fig.~\ref{Fig_S_moduli}. The modulus of ungated split function is evaluated fully analytically, while the analytical approximation (\ref{eq-abs-gated}) to the modulus of gated split function is very accurate at $k_0d\ll 1$ .

\begin{figure}
	\center{\includegraphics[width=0.9\linewidth]{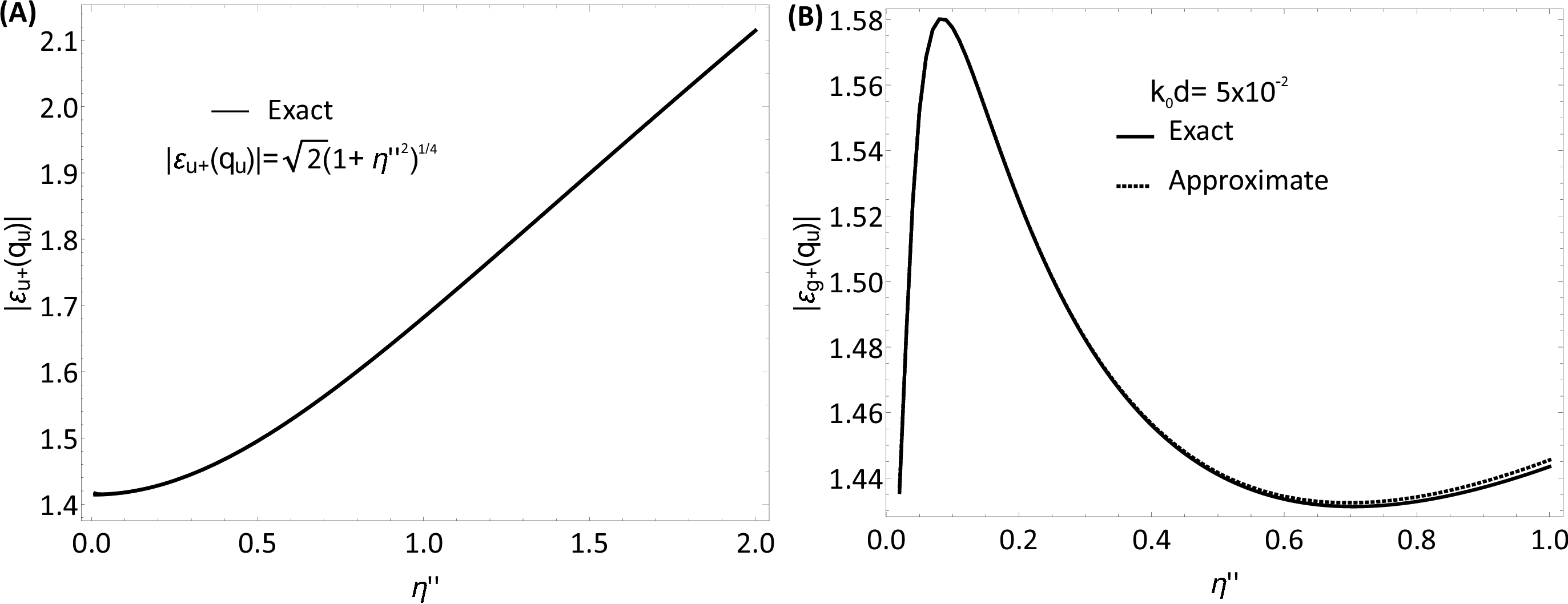}}
	\caption{Absolute values of split functions $\left| {{\varepsilon }_{u+}}({{q}_{u}}) \right|$ and $\left| {{\varepsilon }_{g+}}({{q}_{u}}) \right|$ vs imaginary part of 2DES conductivity $\eta''$. Solid lines show the full numerical results, dashed lines are analytical approximations with eqs. (\ref{eq-abs-ungated}) and (\ref{eq-abs-gated}). For gated split function, the analytical approximation is exact. }
	\label{Fig_S_moduli}
\end{figure}

\subsection{Comparison of various approximations}

We proceed to compare various approximations to the plasmon reflection coefficient $|r|$ and ${\rm arg} r$. At the simplest level, we set the imaginary part of conductivity $\eta'' \rightarrow 0$. This corresponds to the 'ultra-confined limit'. The respective reflectance is denoted as ${{r}^{\left( 1 \right)}}$. In that limit, radiation corrections are fully neglected, and reflectance is approximated by:
\begin{equation}
	\label{eq-ref-ultraconf}
	{{r}^{\left( 1 \right)}}=\left| \frac{1-{{q}_{u}}/{{q}_{g}}}{1+{{q}_{u}}/{{q}_{g}}} \right|\exp \left( -i\frac{\pi }{4}-i{{\phi }_{g}}({{q}_{u}}) \right),
\end{equation}
where the phase due to the gated evanescent fields
\begin{equation}
	{{\phi }_{g}}({{q}_{u}})=-\frac{1}{\pi }\int\limits_{-\infty }^{+\infty }{\frac{d\xi }{\xi -1}\ln \frac{1-|\xi |(1-{{e}^{-2{{q}_{u}}d|\xi |}})}{1-{{\xi }^{2}}/\xi _{g}^{2}({{q}_{u}}d)}}
\end{equation}

The comparison of ultra confined approximation (\ref{eq-ref-ultraconf}) with full numerical result is shown in Fig.~\ref{Fig_S_r_naive}. As expected, the agreement is exact at $\eta''\ll1$, but quicly gets worse with increasing $\eta''$

\begin{figure}
	\center{\includegraphics[width=0.9\linewidth]{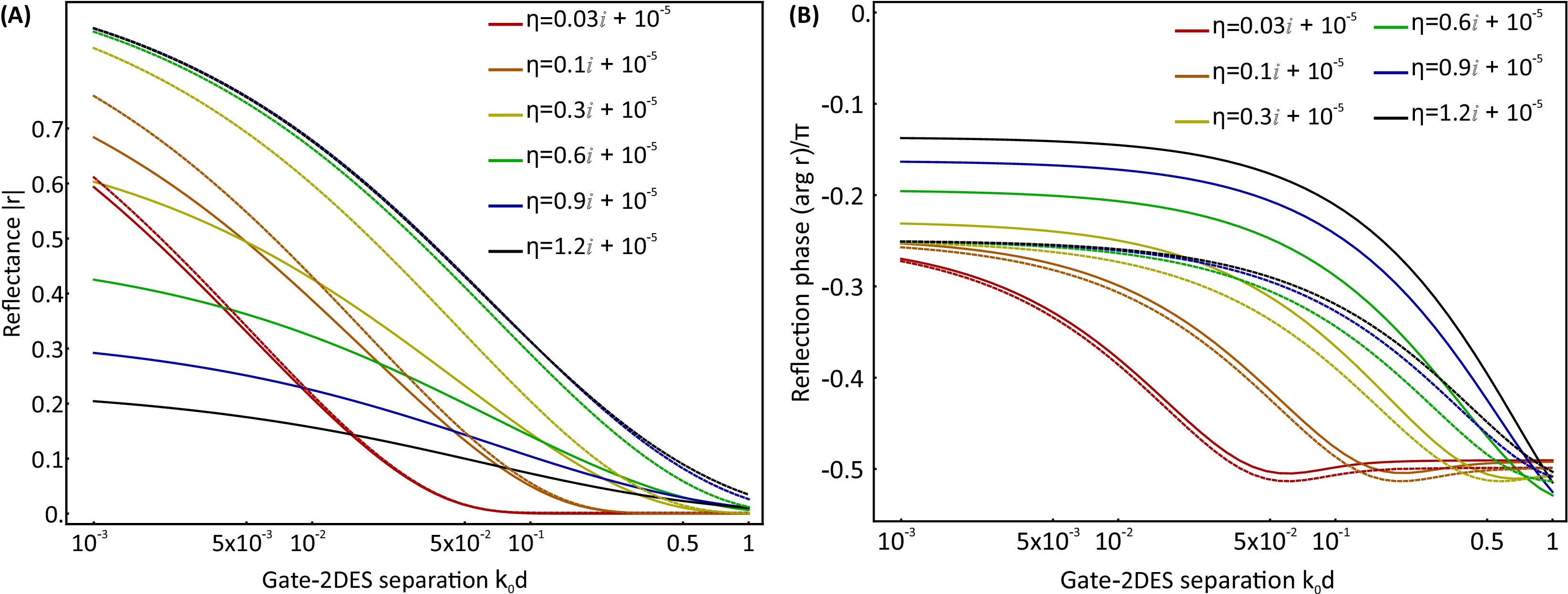}}
	\caption{Comparison of 'ultra-confined approximation' (\ref{eq-ref-ultraconf}) to reflectance (dashed lines) with the full numerical result (solid lines). Panel (A) shows the absolute values of reflectance, panel (B) shows the reflection phase }
	\label{Fig_S_r_naive}
\end{figure}

An advanced approximation to the reflectance includes (1) radiation corrections to the absolute values of split functions, eqs. (\ref{eq-abs-ungated}) and (\ref{eq-abs-gated}) (2) leading-log approximation to the phase of ungated function, eq. \ref{eq-ungated-phase-lead-log}. Collecting these factors, we obtain an advanced approximation $r^{(2)}$:
\begin{equation}
	\label{eq-refl-advanced}
	{{r}^{\left( 2 \right)}}=\left| \frac{1-{{q}_{u}}/{{q}_{g}}}{1+{{q}_{u}}/{{q}_{g}}} \right|\frac{1}{1+{{{{{\eta }'}'}}^{2}}}\frac{1}{{{\eta }'}'+\sqrt{1+{{{{{\eta }'}'}}^{2}}}}\frac{1}{\varepsilon _{g,\text{rad}}^{2}({{q}_{u}})}\exp \left( -i\frac{\pi }{4}-i{{\phi }_{g}}({{q}_{u}})+\frac{i}{\pi }{{{{\eta }''}}^{2}}\ln \frac{1}{{{\eta }''}} \right)
\end{equation}
The comparison of advanced approximation (\ref{eq-refl-advanced}) with full numerical result, eq. (10) of the main text, is shown in Fig.~\ref{Fig_S_r_better}. The approximate absolute values of reflectance $|r^{(2)}|$ are almost indistinguishable from the full numerical result; a tiny difference comes from the approximate nature of eq. (\ref{eq-abs-gated}) for $|\varepsilon_g(q_u)|$. Phases ${\rm arg}r^{(2)}$ match the full numerical result also quite accurately; deviations at $\eta''\sim 1$ are due to the approximate nature of the leading-log approximation (\ref{eq-ungated-phase-lead-log}) to the phase $\phi_u(q_u)$. 

\begin{figure}
	\center{\includegraphics[width=0.9\linewidth]{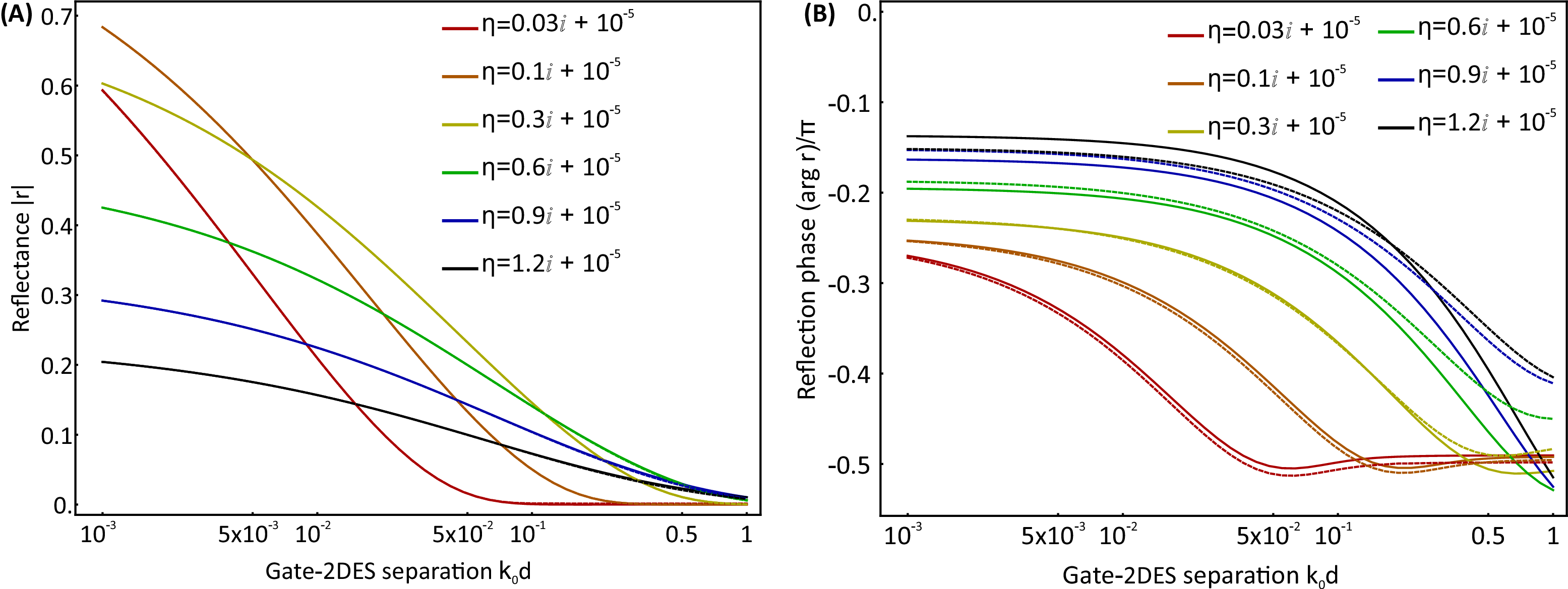}}
	\caption{Comparison of advanced analytical approximation (\ref{eq-ref-ultraconf}) to reflectance (dashed lines) with the full numerical result (solid lines). Panel (A) shows the absolute values of reflectance, panel (B) shows the reflection phase }
	\label{Fig_S_r_better}
\end{figure}

\section{Odd-$n$ slot plasmon modes}
Here we show the possibility of exciting odd gap modes in the case of oblique incidence of the wave on the structure. The field of the obliquely incident wave is non-uniform, which is necessary for the excitation of asymmetric plasmon modes. Such odd modes have a small dipole moment, so their excitation efficiency is usually lower than that of even modes, which is especially evident for large slit widths (\ref{Fig.S4_colormap}).
\begin{figure}
	\center{\includegraphics[width=0.9\linewidth]{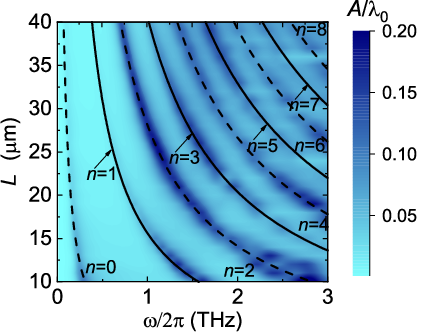}}
	\caption{The spectrum of absorption cross section $A$ normalized at free-space wavelength $\lambda_0$ as function of $L$ in case of oblique incident wave (incidence angle angle $\alpha=\pi/3$). The dashed and solid curves show thee analytically calculated frequencies of the even and odd plasmon modes, respectively.}
	\label{Fig.S4_colormap}
\end{figure}
\end{widetext}

\bibliography{apssamp.bib}

\providecommand{\noopsort}[1]{}\providecommand{\singleletter}[1]{#1}%
\begin{thebibliography}{54}%
\makeatletter
\providecommand \@ifxundefined [1]{%
 \@ifx{#1\undefined}
}%
\providecommand \@ifnum [1]{%
 \ifnum #1\expandafter \@firstoftwo
 \else \expandafter \@secondoftwo
 \fi
}%
\providecommand \@ifx [1]{%
 \ifx #1\expandafter \@firstoftwo
 \else \expandafter \@secondoftwo
 \fi
}%
\providecommand \natexlab [1]{#1}%
\providecommand \enquote  [1]{``#1''}%
\providecommand \bibnamefont  [1]{#1}%
\providecommand \bibfnamefont [1]{#1}%
\providecommand \citenamefont [1]{#1}%
\providecommand \href@noop [0]{\@secondoftwo}%
\providecommand \href [0]{\begingroup \@sanitize@url \@href}%
\providecommand \@href[1]{\@@startlink{#1}\@@href}%
\providecommand \@@href[1]{\endgroup#1\@@endlink}%
\providecommand \@sanitize@url [0]{\catcode `\\12\catcode `\$12\catcode
  `\&12\catcode `\#12\catcode `\^12\catcode `\_12\catcode `\%12\relax}%
\providecommand \@@startlink[1]{}%
\providecommand \@@endlink[0]{}%
\providecommand \url  [0]{\begingroup\@sanitize@url \@url }%
\providecommand \@url [1]{\endgroup\@href {#1}{\urlprefix }}%
\providecommand \urlprefix  [0]{URL }%
\providecommand \Eprint [0]{\href }%
\providecommand \doibase [0]{https://doi.org/}%
\providecommand \selectlanguage [0]{\@gobble}%
\providecommand \bibinfo  [0]{\@secondoftwo}%
\providecommand \bibfield  [0]{\@secondoftwo}%
\providecommand \translation [1]{[#1]}%
\providecommand \BibitemOpen [0]{}%
\providecommand \bibitemStop [0]{}%
\providecommand \bibitemNoStop [0]{.\EOS\space}%
\providecommand \EOS [0]{\spacefactor3000\relax}%
\providecommand \BibitemShut  [1]{\csname bibitem#1\endcsname}%
\let\auto@bib@innerbib\@empty
\bibitem [{\citenamefont {Maier}\ \emph {et~al.}(2007)\citenamefont {Maier}
  \emph {et~al.}}]{Maier:2007}%
  \BibitemOpen
  \bibfield  {author} {\bibinfo {author} {\bibfnamefont {S.~A.}\ \bibnamefont
  {Maier}} \emph {et~al.},\ }\href {https://doi.org/10.1007/0-387-37825-1}
  {\emph {\bibinfo {title} {Plasmonics: fundamentals and applications}}},\
  Vol.~\bibinfo {volume} {1}\ (\bibinfo  {publisher} {Springer},\ \bibinfo
  {year} {2007})\BibitemShut {NoStop}%
\bibitem [{\citenamefont {Barnes}\ \emph {et~al.}(2003)\citenamefont {Barnes},
  \citenamefont {Dereux},\ and\ \citenamefont {Ebbesen}}]{Barnes:2003}%
  \BibitemOpen
  \bibfield  {author} {\bibinfo {author} {\bibfnamefont {W.~L.}\ \bibnamefont
  {Barnes}}, \bibinfo {author} {\bibfnamefont {A.}~\bibnamefont {Dereux}},\
  and\ \bibinfo {author} {\bibfnamefont {T.~W.}\ \bibnamefont {Ebbesen}},\
  }\bibfield  {title} {\bibinfo {title} {Surface plasmon subwavelength
  optics},\ }\href {https://doi.org/10.1038/nature01937} {\bibfield  {journal}
  {\bibinfo  {journal} {Nature}\ }\textbf {\bibinfo {volume} {424}},\ \bibinfo
  {pages} {824} (\bibinfo {year} {2003})}\BibitemShut {NoStop}%
\bibitem [{\citenamefont {Stern}(1967)}]{Stern1967}%
  \BibitemOpen
  \bibfield  {author} {\bibinfo {author} {\bibfnamefont {F.}~\bibnamefont
  {Stern}},\ }\bibfield  {title} {\bibinfo {title} {{Polarizability of a
  Two-Dimensional Electron Gas}},\ }\href
  {https://doi.org/10.1103/PhysRevLett.18.546} {\bibfield  {journal} {\bibinfo
  {journal} {Physical Review Letters}\ }\textbf {\bibinfo {volume} {18}},\
  \bibinfo {pages} {546} (\bibinfo {year} {1967})}\BibitemShut {NoStop}%
\bibitem [{\citenamefont {Woessner}\ \emph {et~al.}(2015)\citenamefont
  {Woessner}, \citenamefont {Lundeberg}, \citenamefont {Gao}, \citenamefont
  {Principi}, \citenamefont {Alonso-Gonz{\'{a}}lez}, \citenamefont {Carrega},
  \citenamefont {Watanabe}, \citenamefont {Taniguchi}, \citenamefont {Vignale},
  \citenamefont {Polini}, \citenamefont {Hone}, \citenamefont {Hillenbrand},\
  and\ \citenamefont {Koppens}}]{Woessner2015}%
  \BibitemOpen
  \bibfield  {author} {\bibinfo {author} {\bibfnamefont {A.}~\bibnamefont
  {Woessner}}, \bibinfo {author} {\bibfnamefont {M.~B.}\ \bibnamefont
  {Lundeberg}}, \bibinfo {author} {\bibfnamefont {Y.}~\bibnamefont {Gao}},
  \bibinfo {author} {\bibfnamefont {A.}~\bibnamefont {Principi}}, \bibinfo
  {author} {\bibfnamefont {P.}~\bibnamefont {Alonso-Gonz{\'{a}}lez}}, \bibinfo
  {author} {\bibfnamefont {M.}~\bibnamefont {Carrega}}, \bibinfo {author}
  {\bibfnamefont {K.}~\bibnamefont {Watanabe}}, \bibinfo {author}
  {\bibfnamefont {T.}~\bibnamefont {Taniguchi}}, \bibinfo {author}
  {\bibfnamefont {G.}~\bibnamefont {Vignale}}, \bibinfo {author} {\bibfnamefont
  {M.}~\bibnamefont {Polini}}, \bibinfo {author} {\bibfnamefont
  {J.}~\bibnamefont {Hone}}, \bibinfo {author} {\bibfnamefont {R.}~\bibnamefont
  {Hillenbrand}},\ and\ \bibinfo {author} {\bibfnamefont {F.~H.~L.}\
  \bibnamefont {Koppens}},\ }\bibfield  {title} {\bibinfo {title} {{Highly
  confined low-loss plasmons in graphene–boron nitride heterostructures}},\
  }\href {https://doi.org/10.1038/nmat4169} {\bibfield  {journal} {\bibinfo
  {journal} {Nature Materials}\ }\textbf {\bibinfo {volume} {14}},\ \bibinfo
  {pages} {421} (\bibinfo {year} {2015})}\BibitemShut {NoStop}%
\bibitem [{\citenamefont {Ni}\ \emph {et~al.}(2018)\citenamefont {Ni},
  \citenamefont {McLeod}, \citenamefont {Sun}, \citenamefont {Wang},
  \citenamefont {Xiong}, \citenamefont {Post}, \citenamefont {Sunku},
  \citenamefont {Jiang}, \citenamefont {Hone}, \citenamefont {Dean},
  \citenamefont {Fogler},\ and\ \citenamefont {Basov}}]{Ni_limits_plasmonics}%
  \BibitemOpen
  \bibfield  {author} {\bibinfo {author} {\bibfnamefont {G.~X.}\ \bibnamefont
  {Ni}}, \bibinfo {author} {\bibfnamefont {A.~S.}\ \bibnamefont {McLeod}},
  \bibinfo {author} {\bibfnamefont {Z.}~\bibnamefont {Sun}}, \bibinfo {author}
  {\bibfnamefont {L.}~\bibnamefont {Wang}}, \bibinfo {author} {\bibfnamefont
  {L.}~\bibnamefont {Xiong}}, \bibinfo {author} {\bibfnamefont {K.~W.}\
  \bibnamefont {Post}}, \bibinfo {author} {\bibfnamefont {S.~S.}\ \bibnamefont
  {Sunku}}, \bibinfo {author} {\bibfnamefont {B.-Y.}\ \bibnamefont {Jiang}},
  \bibinfo {author} {\bibfnamefont {J.}~\bibnamefont {Hone}}, \bibinfo {author}
  {\bibfnamefont {C.~R.}\ \bibnamefont {Dean}}, \bibinfo {author}
  {\bibfnamefont {M.~M.}\ \bibnamefont {Fogler}},\ and\ \bibinfo {author}
  {\bibfnamefont {D.~N.}\ \bibnamefont {Basov}},\ }\bibfield  {title} {\bibinfo
  {title} {{Fundamental limits to graphene plasmonics}},\ }\href
  {https://doi.org/10.1038/s41586-018-0136-9} {\bibfield  {journal} {\bibinfo
  {journal} {Nature}\ }\textbf {\bibinfo {volume} {557}},\ \bibinfo {pages}
  {530} (\bibinfo {year} {2018})}\BibitemShut {NoStop}%
\bibitem [{\citenamefont {Andress}\ \emph {et~al.}(2012)\citenamefont
  {Andress}, \citenamefont {Yoon}, \citenamefont {Yeung}, \citenamefont {Qin},
  \citenamefont {West}, \citenamefont {Pfeiffer},\ and\ \citenamefont
  {Ham}}]{Ham:2012}%
  \BibitemOpen
  \bibfield  {author} {\bibinfo {author} {\bibfnamefont {W.~F.}\ \bibnamefont
  {Andress}}, \bibinfo {author} {\bibfnamefont {H.}~\bibnamefont {Yoon}},
  \bibinfo {author} {\bibfnamefont {K.~Y.}\ \bibnamefont {Yeung}}, \bibinfo
  {author} {\bibfnamefont {L.}~\bibnamefont {Qin}}, \bibinfo {author}
  {\bibfnamefont {K.}~\bibnamefont {West}}, \bibinfo {author} {\bibfnamefont
  {L.}~\bibnamefont {Pfeiffer}},\ and\ \bibinfo {author} {\bibfnamefont
  {D.}~\bibnamefont {Ham}},\ }\bibfield  {title} {\bibinfo {title}
  {Ultra-subwavelength two-dimensional plasmonic circuits},\ }\href
  {https://doi.org/10.1021/nl300046g} {\bibfield  {journal} {\bibinfo
  {journal} {Nano letters}\ }\textbf {\bibinfo {volume} {12}},\ \bibinfo
  {pages} {2272} (\bibinfo {year} {2012})}\BibitemShut {NoStop}%
\bibitem [{\citenamefont {Ritchie}(1957)}]{Ritchie_Plasmons}%
  \BibitemOpen
  \bibfield  {author} {\bibinfo {author} {\bibfnamefont {R.~H.}\ \bibnamefont
  {Ritchie}},\ }\bibfield  {title} {\bibinfo {title} {Plasma losses by fast
  electrons in thin films},\ }\href {https://doi.org/10.1103/PhysRev.106.874}
  {\bibfield  {journal} {\bibinfo  {journal} {Phys. Rev.}\ }\textbf {\bibinfo
  {volume} {106}},\ \bibinfo {pages} {874} (\bibinfo {year}
  {1957})}\BibitemShut {NoStop}%
\bibitem [{\citenamefont {Chaplik}(1972)}]{chaplik1972possible}%
  \BibitemOpen
  \bibfield  {author} {\bibinfo {author} {\bibfnamefont {A.}~\bibnamefont
  {Chaplik}},\ }\bibfield  {title} {\bibinfo {title} {Possible crystallization
  of charge carriers in low-density inversion layers},\ }\href@noop {}
  {\bibfield  {journal} {\bibinfo  {journal} {Soviet Journal of Experimental
  and Theoretical Physics}\ }\textbf {\bibinfo {volume} {35}},\ \bibinfo
  {pages} {395} (\bibinfo {year} {1972})}\BibitemShut {NoStop}%
\bibitem [{\citenamefont {Ryzhii}\ \emph {et~al.}(2007)\citenamefont {Ryzhii},
  \citenamefont {Satou},\ and\ \citenamefont {Otsuji}}]{Ryzhii2007a}%
  \BibitemOpen
  \bibfield  {author} {\bibinfo {author} {\bibfnamefont {V.}~\bibnamefont
  {Ryzhii}}, \bibinfo {author} {\bibfnamefont {A.}~\bibnamefont {Satou}},\ and\
  \bibinfo {author} {\bibfnamefont {T.}~\bibnamefont {Otsuji}},\ }\bibfield
  {title} {\bibinfo {title} {{Plasma waves in two-dimensional electron-hole
  system in gated graphene heterostructures}},\ }\href
  {https://doi.org/10.1063/1.2426904} {\bibfield  {journal} {\bibinfo
  {journal} {Journal of Applied Physics}\ }\textbf {\bibinfo {volume} {101}},\
  \bibinfo {pages} {024509} (\bibinfo {year} {2007})}\BibitemShut {NoStop}%
\bibitem [{\citenamefont {Woessner}\ \emph {et~al.}(2017)\citenamefont
  {Woessner}, \citenamefont {Gao}, \citenamefont {Torre}, \citenamefont
  {Lundeberg}, \citenamefont {Tan}, \citenamefont {Watanabe}, \citenamefont
  {Taniguchi}, \citenamefont {Hillenbrand}, \citenamefont {Hone}, \citenamefont
  {Polini},\ and\ \citenamefont {Koppens}}]{Woessner_phase_shifter}%
  \BibitemOpen
  \bibfield  {author} {\bibinfo {author} {\bibfnamefont {A.}~\bibnamefont
  {Woessner}}, \bibinfo {author} {\bibfnamefont {Y.}~\bibnamefont {Gao}},
  \bibinfo {author} {\bibfnamefont {I.}~\bibnamefont {Torre}}, \bibinfo
  {author} {\bibfnamefont {M.~B.}\ \bibnamefont {Lundeberg}}, \bibinfo {author}
  {\bibfnamefont {C.}~\bibnamefont {Tan}}, \bibinfo {author} {\bibfnamefont
  {K.}~\bibnamefont {Watanabe}}, \bibinfo {author} {\bibfnamefont
  {T.}~\bibnamefont {Taniguchi}}, \bibinfo {author} {\bibfnamefont
  {R.}~\bibnamefont {Hillenbrand}}, \bibinfo {author} {\bibfnamefont
  {J.}~\bibnamefont {Hone}}, \bibinfo {author} {\bibfnamefont {M.}~\bibnamefont
  {Polini}},\ and\ \bibinfo {author} {\bibfnamefont {F.~H.}\ \bibnamefont
  {Koppens}},\ }\bibfield  {title} {\bibinfo {title} {{Electrical 2$\pi$ phase
  control of infrared light in a 350-nm footprint using graphene plasmons}},\
  }\href {https://doi.org/10.1038/nphoton.2017.98} {\bibfield  {journal}
  {\bibinfo  {journal} {Nature Photonics}\ }\textbf {\bibinfo {volume} {11}},\
  \bibinfo {pages} {421} (\bibinfo {year} {2017})}\BibitemShut {NoStop}%
\bibitem [{\citenamefont {Alonso-Gonzalez}\ \emph {et~al.}(2014)\citenamefont
  {Alonso-Gonzalez}, \citenamefont {Nikitin}, \citenamefont {Golmar},
  \citenamefont {Centeno}, \citenamefont {Pesquera}, \citenamefont {Velez},
  \citenamefont {Chen}, \citenamefont {Navickaite}, \citenamefont {Koppens},
  \citenamefont {Zurutuza}, \citenamefont {Casanova}, \citenamefont {Hueso},\
  and\ \citenamefont {Hillenbrand}}]{Alonso-Gonzalez_antenna_launching}%
  \BibitemOpen
  \bibfield  {author} {\bibinfo {author} {\bibfnamefont {P.}~\bibnamefont
  {Alonso-Gonzalez}}, \bibinfo {author} {\bibfnamefont {A.~Y.}\ \bibnamefont
  {Nikitin}}, \bibinfo {author} {\bibfnamefont {F.}~\bibnamefont {Golmar}},
  \bibinfo {author} {\bibfnamefont {A.}~\bibnamefont {Centeno}}, \bibinfo
  {author} {\bibfnamefont {A.}~\bibnamefont {Pesquera}}, \bibinfo {author}
  {\bibfnamefont {S.}~\bibnamefont {Velez}}, \bibinfo {author} {\bibfnamefont
  {J.}~\bibnamefont {Chen}}, \bibinfo {author} {\bibfnamefont {G.}~\bibnamefont
  {Navickaite}}, \bibinfo {author} {\bibfnamefont {F.}~\bibnamefont {Koppens}},
  \bibinfo {author} {\bibfnamefont {A.}~\bibnamefont {Zurutuza}}, \bibinfo
  {author} {\bibfnamefont {F.}~\bibnamefont {Casanova}}, \bibinfo {author}
  {\bibfnamefont {L.~E.}\ \bibnamefont {Hueso}},\ and\ \bibinfo {author}
  {\bibfnamefont {R.}~\bibnamefont {Hillenbrand}},\ }\bibfield  {title}
  {\bibinfo {title} {{Controlling graphene plasmons with resonant metal
  antennas and spatial conductivity patterns}},\ }\href
  {https://doi.org/10.1126/science.1253202} {\bibfield  {journal} {\bibinfo
  {journal} {Science}\ }\textbf {\bibinfo {volume} {344}},\ \bibinfo {pages}
  {1369} (\bibinfo {year} {2014})}\BibitemShut {NoStop}%
\bibitem [{\citenamefont {Dias}\ and\ \citenamefont {{Garc{\'{i}}a De
  Abajo}}(2019)}]{Abajo_Limits}%
  \BibitemOpen
  \bibfield  {author} {\bibinfo {author} {\bibfnamefont {E.~J.}\ \bibnamefont
  {Dias}}\ and\ \bibinfo {author} {\bibfnamefont {F.~J.}\ \bibnamefont
  {{Garc{\'{i}}a De Abajo}}},\ }\bibfield  {title} {\bibinfo {title}
  {{Fundamental Limits to the Coupling between Light and 2D Polaritons by Small
  Scatterers}},\ }\href {https://doi.org/10.1021/acsnano.8b09283} {\bibfield
  {journal} {\bibinfo  {journal} {ACS Nano}\ }\textbf {\bibinfo {volume}
  {13}},\ \bibinfo {pages} {5184} (\bibinfo {year} {2019})}\BibitemShut
  {NoStop}%
\bibitem [{\citenamefont {Maradudin}\ \emph {et~al.}(1990)\citenamefont
  {Maradudin}, \citenamefont {Wallis},\ and\ \citenamefont
  {Stegeman}}]{Maradudin1990}%
  \BibitemOpen
  \bibfield  {author} {\bibinfo {author} {\bibfnamefont {A.}~\bibnamefont
  {Maradudin}}, \bibinfo {author} {\bibfnamefont {R.}~\bibnamefont {Wallis}},\
  and\ \bibinfo {author} {\bibfnamefont {G.}~\bibnamefont {Stegeman}},\
  }\bibfield  {title} {\bibinfo {title} {{The optics of surface and guided wave
  polaritons}},\ }\href {https://doi.org/10.1016/0079-6816(90)90004-4}
  {\bibfield  {journal} {\bibinfo  {journal} {Progress in Surface Science}\
  }\textbf {\bibinfo {volume} {33}},\ \bibinfo {pages} {171} (\bibinfo {year}
  {1990})}\BibitemShut {NoStop}%
\bibitem [{\citenamefont {Agranovich}\ \emph {et~al.}(1981)\citenamefont
  {Agranovich}, \citenamefont {Kravtsov},\ and\ \citenamefont
  {Leskova}}]{agranovich1981diffraction}%
  \BibitemOpen
  \bibfield  {author} {\bibinfo {author} {\bibfnamefont {V.}~\bibnamefont
  {Agranovich}}, \bibinfo {author} {\bibfnamefont {V.}~\bibnamefont
  {Kravtsov}},\ and\ \bibinfo {author} {\bibfnamefont {T.}~\bibnamefont
  {Leskova}},\ }\bibfield  {title} {\bibinfo {title} {Diffraction of the
  surface polaritons by an impedance step in the region of resonance with
  oscillations in a transition layer},\ }\href@noop {} {\bibfield  {journal}
  {\bibinfo  {journal} {Zh. Eksp. Teor. Fiz.}\ }\textbf {\bibinfo {volume}
  {81}},\ \bibinfo {pages} {1828} (\bibinfo {year} {1981})}\BibitemShut
  {NoStop}%
\bibitem [{\citenamefont {Pincemin}\ \emph {et~al.}(1994)\citenamefont
  {Pincemin}, \citenamefont {Maradudin}, \citenamefont {Boardman},\ and\
  \citenamefont {Greffet}}]{Pincemin1994}%
  \BibitemOpen
  \bibfield  {author} {\bibinfo {author} {\bibfnamefont {F.}~\bibnamefont
  {Pincemin}}, \bibinfo {author} {\bibfnamefont {A.~A.}\ \bibnamefont
  {Maradudin}}, \bibinfo {author} {\bibfnamefont {A.~D.}\ \bibnamefont
  {Boardman}},\ and\ \bibinfo {author} {\bibfnamefont {J.-J.}\ \bibnamefont
  {Greffet}},\ }\bibfield  {title} {\bibinfo {title} {{Scattering of a surface
  plasmon polariton by a surface defect}},\ }\href
  {https://doi.org/10.1103/PhysRevB.50.15261} {\bibfield  {journal} {\bibinfo
  {journal} {Physical Review B}\ }\textbf {\bibinfo {volume} {50}},\ \bibinfo
  {pages} {15261} (\bibinfo {year} {1994})}\BibitemShut {NoStop}%
\bibitem [{\citenamefont {Ditlbacher}\ \emph {et~al.}(2002)\citenamefont
  {Ditlbacher}, \citenamefont {Krenn}, \citenamefont {Schider}, \citenamefont
  {Leitner},\ and\ \citenamefont {Aussenegg}}]{Ditlbacher2002}%
  \BibitemOpen
  \bibfield  {author} {\bibinfo {author} {\bibfnamefont {H.}~\bibnamefont
  {Ditlbacher}}, \bibinfo {author} {\bibfnamefont {J.~R.}\ \bibnamefont
  {Krenn}}, \bibinfo {author} {\bibfnamefont {G.}~\bibnamefont {Schider}},
  \bibinfo {author} {\bibfnamefont {A.}~\bibnamefont {Leitner}},\ and\ \bibinfo
  {author} {\bibfnamefont {F.~R.}\ \bibnamefont {Aussenegg}},\ }\bibfield
  {title} {\bibinfo {title} {{Two-dimensional optics with surface plasmon
  polaritons}},\ }\href {https://doi.org/10.1063/1.1506018} {\bibfield
  {journal} {\bibinfo  {journal} {Applied Physics Letters}\ }\textbf {\bibinfo
  {volume} {81}},\ \bibinfo {pages} {1762} (\bibinfo {year}
  {2002})}\BibitemShut {NoStop}%
\bibitem [{\citenamefont {Malyuzhinets}(1958)}]{malyuzhinets1958excitation}%
  \BibitemOpen
  \bibfield  {author} {\bibinfo {author} {\bibfnamefont {G.}~\bibnamefont
  {Malyuzhinets}},\ }\bibfield  {title} {\bibinfo {title} {The excitation,
  reflection and radiation of surface waves in a wedge-like region with given
  face impedances},\ }\href
  {https://www.mathnet.ru/php/archive.phtml?wshow=paper&jrnid=dan&paperid=23284&option_lang=rus}
  {\bibfield  {journal} {\bibinfo  {journal} {Doklady Akademii Nauk}\ }\textbf
  {\bibinfo {volume} {121}},\ \bibinfo {pages} {436} (\bibinfo {year}
  {1958})}\BibitemShut {NoStop}%
\bibitem [{\citenamefont {Sommerfeld}(1896)}]{Sommerfeld:1896}%
  \BibitemOpen
  \bibfield  {author} {\bibinfo {author} {\bibfnamefont {A.}~\bibnamefont
  {Sommerfeld}},\ }\bibfield  {title} {\bibinfo {title} {Mathematische theorie
  der diffraction},\ }\href {https://doi.org/10.1007/978-0-8176-8196-8}
  {\bibfield  {journal} {\bibinfo  {journal} {Mathematische Annalen}\ }\textbf
  {\bibinfo {volume} {16}},\ \bibinfo {pages} {317} (\bibinfo {year}
  {1896})}\BibitemShut {NoStop}%
\bibitem [{\citenamefont {Nikitin}\ \emph {et~al.}(2014)\citenamefont
  {Nikitin}, \citenamefont {Low},\ and\ \citenamefont
  {Martin-Moreno}}]{Moreno_phase}%
  \BibitemOpen
  \bibfield  {author} {\bibinfo {author} {\bibfnamefont {A.~Y.}\ \bibnamefont
  {Nikitin}}, \bibinfo {author} {\bibfnamefont {T.}~\bibnamefont {Low}},\ and\
  \bibinfo {author} {\bibfnamefont {L.}~\bibnamefont {Martin-Moreno}},\
  }\bibfield  {title} {\bibinfo {title} {Anomalous reflection phase of graphene
  plasmons and its influence on resonators},\ }\href
  {https://doi.org/10.1103/PhysRevB.90.041407} {\bibfield  {journal} {\bibinfo
  {journal} {Phys. Rev. B}\ }\textbf {\bibinfo {volume} {90}},\ \bibinfo
  {pages} {041407} (\bibinfo {year} {2014})}\BibitemShut {NoStop}%
\bibitem [{\citenamefont {Rejaei}\ and\ \citenamefont
  {Khavasi}(2015)}]{Khavasi_pi4_phase}%
  \BibitemOpen
  \bibfield  {author} {\bibinfo {author} {\bibfnamefont {B.}~\bibnamefont
  {Rejaei}}\ and\ \bibinfo {author} {\bibfnamefont {A.}~\bibnamefont
  {Khavasi}},\ }\bibfield  {title} {\bibinfo {title} {{Scattering of surface
  plasmons on graphene by a discontinuity in surface conductivity}},\ }\href
  {https://doi.org/10.1088/2040-8978/17/7/075002} {\bibfield  {journal}
  {\bibinfo  {journal} {Journal of Optics (United Kingdom)}\ }\textbf {\bibinfo
  {volume} {17}},\ \bibinfo {pages} {75002} (\bibinfo {year}
  {2015})}\BibitemShut {NoStop}%
\bibitem [{\citenamefont {Svintsov}\ and\ \citenamefont
  {Alymov}(2023)}]{Alymov_Refraction}%
  \BibitemOpen
  \bibfield  {author} {\bibinfo {author} {\bibfnamefont {D.~A.}\ \bibnamefont
  {Svintsov}}\ and\ \bibinfo {author} {\bibfnamefont {G.~V.}\ \bibnamefont
  {Alymov}},\ }\bibfield  {title} {\bibinfo {title} {{Refraction laws for
  two-dimensional plasmons}},\ }\href
  {https://doi.org/10.1103/PhysRevB.108.L121410} {\bibfield  {journal}
  {\bibinfo  {journal} {Physical Review B}\ }\textbf {\bibinfo {volume}
  {108}},\ \bibinfo {pages} {L121410} (\bibinfo {year} {2023})}\BibitemShut
  {NoStop}%
\bibitem [{\citenamefont {Jiang}\ \emph {et~al.}(2018)\citenamefont {Jiang},
  \citenamefont {Mele},\ and\ \citenamefont {Fogler}}]{Fogler_1d_junction}%
  \BibitemOpen
  \bibfield  {author} {\bibinfo {author} {\bibfnamefont {B.-Y.}\ \bibnamefont
  {Jiang}}, \bibinfo {author} {\bibfnamefont {E.~J.}\ \bibnamefont {Mele}},\
  and\ \bibinfo {author} {\bibfnamefont {M.~M.}\ \bibnamefont {Fogler}},\
  }\bibfield  {title} {\bibinfo {title} {{Theory of plasmon reflection by a 1D
  junction}},\ }\href {https://doi.org/10.1364/oe.26.017209} {\bibfield
  {journal} {\bibinfo  {journal} {Optics Express}\ }\textbf {\bibinfo {volume}
  {26}},\ \bibinfo {pages} {17209} (\bibinfo {year} {2018})}\BibitemShut
  {NoStop}%
\bibitem [{\citenamefont {Semenenko}\ \emph {et~al.}(2020)\citenamefont
  {Semenenko}, \citenamefont {Liu},\ and\ \citenamefont
  {Perebeinos}}]{Semenenko_scattering}%
  \BibitemOpen
  \bibfield  {author} {\bibinfo {author} {\bibfnamefont {V.}~\bibnamefont
  {Semenenko}}, \bibinfo {author} {\bibfnamefont {M.}~\bibnamefont {Liu}},\
  and\ \bibinfo {author} {\bibfnamefont {V.}~\bibnamefont {Perebeinos}},\
  }\bibfield  {title} {\bibinfo {title} {Scattering of quasistatic plasmons
  from one-dimensional junctions of graphene: Transfer matrices, fresnel
  relations, and nonlocality},\ }\href
  {https://doi.org/10.1103/PhysRevApplied.14.024049} {\bibfield  {journal}
  {\bibinfo  {journal} {Phys. Rev. Appl.}\ }\textbf {\bibinfo {volume} {14}},\
  \bibinfo {pages} {024049} (\bibinfo {year} {2020})}\BibitemShut {NoStop}%
\bibitem [{\citenamefont {Jadidi}\ \emph {et~al.}(2015)\citenamefont {Jadidi},
  \citenamefont {Sushkov}, \citenamefont {Myers-Ward}, \citenamefont {Boyd},
  \citenamefont {Daniels}, \citenamefont {Gaskill}, \citenamefont {Fuhrer},
  \citenamefont {Drew},\ and\ \citenamefont {Murphy}}]{Jadidi2015a}%
  \BibitemOpen
  \bibfield  {author} {\bibinfo {author} {\bibfnamefont {M.~M.}\ \bibnamefont
  {Jadidi}}, \bibinfo {author} {\bibfnamefont {A.~B.}\ \bibnamefont {Sushkov}},
  \bibinfo {author} {\bibfnamefont {R.~L.}\ \bibnamefont {Myers-Ward}},
  \bibinfo {author} {\bibfnamefont {A.~K.}\ \bibnamefont {Boyd}}, \bibinfo
  {author} {\bibfnamefont {K.~M.}\ \bibnamefont {Daniels}}, \bibinfo {author}
  {\bibfnamefont {D.~K.}\ \bibnamefont {Gaskill}}, \bibinfo {author}
  {\bibfnamefont {M.~S.}\ \bibnamefont {Fuhrer}}, \bibinfo {author}
  {\bibfnamefont {H.~D.}\ \bibnamefont {Drew}},\ and\ \bibinfo {author}
  {\bibfnamefont {T.~E.}\ \bibnamefont {Murphy}},\ }\bibfield  {title}
  {\bibinfo {title} {{Tunable Terahertz Hybrid Metal-Graphene Plasmons}},\
  }\href {https://doi.org/10.1021/acs.nanolett.5b03191} {\bibfield  {journal}
  {\bibinfo  {journal} {Nano Letters}\ }\textbf {\bibinfo {volume} {15}},\
  \bibinfo {pages} {7099} (\bibinfo {year} {2015})}\BibitemShut {NoStop}%
\bibitem [{\citenamefont {Tamagnone}\ \emph {et~al.}(2018)\citenamefont
  {Tamagnone}, \citenamefont {Ambrosio}, \citenamefont {Chaudhary},
  \citenamefont {Jauregui}, \citenamefont {Kim}, \citenamefont {Wilson},\ and\
  \citenamefont {Capasso}}]{Tamagnone2018}%
  \BibitemOpen
  \bibfield  {author} {\bibinfo {author} {\bibfnamefont {M.}~\bibnamefont
  {Tamagnone}}, \bibinfo {author} {\bibfnamefont {A.}~\bibnamefont {Ambrosio}},
  \bibinfo {author} {\bibfnamefont {K.}~\bibnamefont {Chaudhary}}, \bibinfo
  {author} {\bibfnamefont {L.~A.}\ \bibnamefont {Jauregui}}, \bibinfo {author}
  {\bibfnamefont {P.}~\bibnamefont {Kim}}, \bibinfo {author} {\bibfnamefont
  {W.~L.}\ \bibnamefont {Wilson}},\ and\ \bibinfo {author} {\bibfnamefont
  {F.}~\bibnamefont {Capasso}},\ }\bibfield  {title} {\bibinfo {title}
  {{Ultra-confined mid-infrared resonant phonon polaritons in van der Waals
  nanostructures}},\ }\href {https://doi.org/10.1126/sciadv.aat7189} {\bibfield
   {journal} {\bibinfo  {journal} {Science Advances}\ }\textbf {\bibinfo
  {volume} {4}},\ \bibinfo {pages} {4} (\bibinfo {year} {2018})}\BibitemShut
  {NoStop}%
\bibitem [{\citenamefont {Olbrich}\ \emph {et~al.}(2016)\citenamefont
  {Olbrich}, \citenamefont {Kamann}, \citenamefont {K{\"{o}}nig}, \citenamefont
  {Munzert}, \citenamefont {Tutsch}, \citenamefont {Eroms}, \citenamefont
  {Weiss}, \citenamefont {Liu}, \citenamefont {Golub}, \citenamefont
  {Ivchenko}, \citenamefont {Popov}, \citenamefont {Fateev}, \citenamefont
  {Mashinsky}, \citenamefont {Fromm}, \citenamefont {Seyller},\ and\
  \citenamefont {Ganichev}}]{Olbrich2016}%
  \BibitemOpen
  \bibfield  {author} {\bibinfo {author} {\bibfnamefont {P.}~\bibnamefont
  {Olbrich}}, \bibinfo {author} {\bibfnamefont {J.}~\bibnamefont {Kamann}},
  \bibinfo {author} {\bibfnamefont {M.}~\bibnamefont {K{\"{o}}nig}}, \bibinfo
  {author} {\bibfnamefont {J.}~\bibnamefont {Munzert}}, \bibinfo {author}
  {\bibfnamefont {L.}~\bibnamefont {Tutsch}}, \bibinfo {author} {\bibfnamefont
  {J.}~\bibnamefont {Eroms}}, \bibinfo {author} {\bibfnamefont
  {D.}~\bibnamefont {Weiss}}, \bibinfo {author} {\bibfnamefont {M.-H.}\
  \bibnamefont {Liu}}, \bibinfo {author} {\bibfnamefont {L.~E.}\ \bibnamefont
  {Golub}}, \bibinfo {author} {\bibfnamefont {E.~L.}\ \bibnamefont {Ivchenko}},
  \bibinfo {author} {\bibfnamefont {V.~V.}\ \bibnamefont {Popov}}, \bibinfo
  {author} {\bibfnamefont {D.~V.}\ \bibnamefont {Fateev}}, \bibinfo {author}
  {\bibfnamefont {K.~V.}\ \bibnamefont {Mashinsky}}, \bibinfo {author}
  {\bibfnamefont {F.}~\bibnamefont {Fromm}}, \bibinfo {author} {\bibfnamefont
  {T.}~\bibnamefont {Seyller}},\ and\ \bibinfo {author} {\bibfnamefont {S.~D.}\
  \bibnamefont {Ganichev}},\ }\bibfield  {title} {\bibinfo {title} {{Terahertz
  ratchet effects in graphene with a lateral superlattice}},\ }\href
  {https://doi.org/10.1103/PhysRevB.93.075422} {\bibfield  {journal} {\bibinfo
  {journal} {Physical Review B}\ }\textbf {\bibinfo {volume} {93}},\ \bibinfo
  {pages} {075422} (\bibinfo {year} {2016})}\BibitemShut {NoStop}%
\bibitem [{\citenamefont {Popov}\ \emph {et~al.}(2011)\citenamefont {Popov},
  \citenamefont {Fateev}, \citenamefont {Otsuji}, \citenamefont {Meziani},
  \citenamefont {Coquillat},\ and\ \citenamefont {Knap}}]{Otsuji:2011}%
  \BibitemOpen
  \bibfield  {author} {\bibinfo {author} {\bibfnamefont {V.~V.}\ \bibnamefont
  {Popov}}, \bibinfo {author} {\bibfnamefont {D.~V.}\ \bibnamefont {Fateev}},
  \bibinfo {author} {\bibfnamefont {T.}~\bibnamefont {Otsuji}}, \bibinfo
  {author} {\bibfnamefont {Y.~M.}\ \bibnamefont {Meziani}}, \bibinfo {author}
  {\bibfnamefont {D.}~\bibnamefont {Coquillat}},\ and\ \bibinfo {author}
  {\bibfnamefont {W.}~\bibnamefont {Knap}},\ }\bibfield  {title} {\bibinfo
  {title} {Plasmonic terahertz detection by a double-grating-gate field-effect
  transistor structure with an asymmetric unit cell},\ }\href
  {https://doi.org/10.1063/1.3670321} {\bibfield  {journal} {\bibinfo
  {journal} {Applied Physics Letters}\ }\textbf {\bibinfo {volume} {99}},\
  \bibinfo {pages} {243504} (\bibinfo {year} {2011})}\BibitemShut {NoStop}%
\bibitem [{\citenamefont {Kachorovskii}\ and\ \citenamefont
  {Shur}(2012)}]{Kachorovskii_gated_instability}%
  \BibitemOpen
  \bibfield  {author} {\bibinfo {author} {\bibfnamefont {V.~Y.}\ \bibnamefont
  {Kachorovskii}}\ and\ \bibinfo {author} {\bibfnamefont {M.~S.}\ \bibnamefont
  {Shur}},\ }\bibfield  {title} {\bibinfo {title} {Current-induced terahertz
  oscillations in plasmonic crystal},\ }\href
  {https://doi.org/10.1063/1.4726273} {\bibfield  {journal} {\bibinfo
  {journal} {Applied Physics Letters}\ }\textbf {\bibinfo {volume} {100}},\
  \bibinfo {pages} {232108} (\bibinfo {year} {2012})}\BibitemShut {NoStop}%
\bibitem [{\citenamefont {Boubanga-Tombet}\ \emph {et~al.}(2020)\citenamefont
  {Boubanga-Tombet}, \citenamefont {Knap}, \citenamefont {Yadav}, \citenamefont
  {Satou}, \citenamefont {But}, \citenamefont {Popov}, \citenamefont
  {Gorbenko}, \citenamefont {Kachorovskii},\ and\ \citenamefont
  {Otsuji}}]{Boubanga-Tombet2020}%
  \BibitemOpen
  \bibfield  {author} {\bibinfo {author} {\bibfnamefont {S.}~\bibnamefont
  {Boubanga-Tombet}}, \bibinfo {author} {\bibfnamefont {W.}~\bibnamefont
  {Knap}}, \bibinfo {author} {\bibfnamefont {D.}~\bibnamefont {Yadav}},
  \bibinfo {author} {\bibfnamefont {A.}~\bibnamefont {Satou}}, \bibinfo
  {author} {\bibfnamefont {D.~B.}\ \bibnamefont {But}}, \bibinfo {author}
  {\bibfnamefont {V.~V.}\ \bibnamefont {Popov}}, \bibinfo {author}
  {\bibfnamefont {I.~V.}\ \bibnamefont {Gorbenko}}, \bibinfo {author}
  {\bibfnamefont {V.}~\bibnamefont {Kachorovskii}},\ and\ \bibinfo {author}
  {\bibfnamefont {T.}~\bibnamefont {Otsuji}},\ }\bibfield  {title} {\bibinfo
  {title} {{Room-Temperature Amplification of Terahertz Radiation by
  Grating-Gate Graphene Structures}},\ }\href
  {https://doi.org/10.1103/PhysRevX.10.031004} {\bibfield  {journal} {\bibinfo
  {journal} {Physical Review X}\ }\textbf {\bibinfo {volume} {10}},\ \bibinfo
  {pages} {031004} (\bibinfo {year} {2020})}\BibitemShut {NoStop}%
\bibitem [{\citenamefont {Xu}\ \emph {et~al.}(2019)\citenamefont {Xu},
  \citenamefont {Xie}, \citenamefont {Zhu}, \citenamefont {Tang}, \citenamefont
  {Singh}, \citenamefont {Wang}, \citenamefont {Ma}, \citenamefont {Chen},\
  and\ \citenamefont {Ying}}]{Xu_Sensor}%
  \BibitemOpen
  \bibfield  {author} {\bibinfo {author} {\bibfnamefont {W.}~\bibnamefont
  {Xu}}, \bibinfo {author} {\bibfnamefont {L.}~\bibnamefont {Xie}}, \bibinfo
  {author} {\bibfnamefont {J.}~\bibnamefont {Zhu}}, \bibinfo {author}
  {\bibfnamefont {L.}~\bibnamefont {Tang}}, \bibinfo {author} {\bibfnamefont
  {R.}~\bibnamefont {Singh}}, \bibinfo {author} {\bibfnamefont
  {C.}~\bibnamefont {Wang}}, \bibinfo {author} {\bibfnamefont {Y.}~\bibnamefont
  {Ma}}, \bibinfo {author} {\bibfnamefont {H.-T.}\ \bibnamefont {Chen}},\ and\
  \bibinfo {author} {\bibfnamefont {Y.}~\bibnamefont {Ying}},\ }\bibfield
  {title} {\bibinfo {title} {Terahertz biosensing with a graphene-metamaterial
  heterostructure platform},\ }\href
  {https://doi.org/https://doi.org/10.1016/j.carbon.2018.09.050} {\bibfield
  {journal} {\bibinfo  {journal} {Carbon}\ }\textbf {\bibinfo {volume} {141}},\
  \bibinfo {pages} {247} (\bibinfo {year} {2019})}\BibitemShut {NoStop}%
\bibitem [{\citenamefont {Shen}\ \emph {et~al.}(2022)\citenamefont {Shen},
  \citenamefont {Liu}, \citenamefont {Shen}, \citenamefont {Qu}, \citenamefont
  {Pickwell-MacPherson}, \citenamefont {Wei},\ and\ \citenamefont
  {Sun}}]{Metasurface_sensor2}%
  \BibitemOpen
  \bibfield  {author} {\bibinfo {author} {\bibfnamefont {S.}~\bibnamefont
  {Shen}}, \bibinfo {author} {\bibfnamefont {X.}~\bibnamefont {Liu}}, \bibinfo
  {author} {\bibfnamefont {Y.}~\bibnamefont {Shen}}, \bibinfo {author}
  {\bibfnamefont {J.}~\bibnamefont {Qu}}, \bibinfo {author} {\bibfnamefont
  {E.}~\bibnamefont {Pickwell-MacPherson}}, \bibinfo {author} {\bibfnamefont
  {X.}~\bibnamefont {Wei}},\ and\ \bibinfo {author} {\bibfnamefont
  {Y.}~\bibnamefont {Sun}},\ }\bibfield  {title} {\bibinfo {title} {Recent
  advances in the development of materials for terahertz metamaterial
  sensing},\ }\href {https://doi.org/https://doi.org/10.1002/adom.202101008}
  {\bibfield  {journal} {\bibinfo  {journal} {Advanced Optical Materials}\
  }\textbf {\bibinfo {volume} {10}},\ \bibinfo {pages} {2101008} (\bibinfo
  {year} {2022})}\BibitemShut {NoStop}%
\bibitem [{\citenamefont {{Alonso Calafell}}\ \emph {et~al.}(2021)\citenamefont
  {{Alonso Calafell}}, \citenamefont {Rozema}, \citenamefont {{Alcaraz
  Iranzo}}, \citenamefont {Trenti}, \citenamefont {Jenke}, \citenamefont {Cox},
  \citenamefont {Kumar}, \citenamefont {Bieliaiev}, \citenamefont {Nanot},
  \citenamefont {Peng}, \citenamefont {Efetov}, \citenamefont {Hong},
  \citenamefont {Kong}, \citenamefont {Englund}, \citenamefont {{Garc{\'{i}}a
  de Abajo}}, \citenamefont {Koppens},\ and\ \citenamefont
  {Walther}}]{AlonsoCalafell2021}%
  \BibitemOpen
  \bibfield  {author} {\bibinfo {author} {\bibfnamefont {I.}~\bibnamefont
  {{Alonso Calafell}}}, \bibinfo {author} {\bibfnamefont {L.~A.}\ \bibnamefont
  {Rozema}}, \bibinfo {author} {\bibfnamefont {D.}~\bibnamefont {{Alcaraz
  Iranzo}}}, \bibinfo {author} {\bibfnamefont {A.}~\bibnamefont {Trenti}},
  \bibinfo {author} {\bibfnamefont {P.~K.}\ \bibnamefont {Jenke}}, \bibinfo
  {author} {\bibfnamefont {J.~D.}\ \bibnamefont {Cox}}, \bibinfo {author}
  {\bibfnamefont {A.}~\bibnamefont {Kumar}}, \bibinfo {author} {\bibfnamefont
  {H.}~\bibnamefont {Bieliaiev}}, \bibinfo {author} {\bibfnamefont
  {S.}~\bibnamefont {Nanot}}, \bibinfo {author} {\bibfnamefont
  {C.}~\bibnamefont {Peng}}, \bibinfo {author} {\bibfnamefont {D.~K.}\
  \bibnamefont {Efetov}}, \bibinfo {author} {\bibfnamefont {J.-Y.}\
  \bibnamefont {Hong}}, \bibinfo {author} {\bibfnamefont {J.}~\bibnamefont
  {Kong}}, \bibinfo {author} {\bibfnamefont {D.~R.}\ \bibnamefont {Englund}},
  \bibinfo {author} {\bibfnamefont {F.~J.}\ \bibnamefont {{Garc{\'{i}}a de
  Abajo}}}, \bibinfo {author} {\bibfnamefont {F.~H.~L.}\ \bibnamefont
  {Koppens}},\ and\ \bibinfo {author} {\bibfnamefont {P.}~\bibnamefont
  {Walther}},\ }\bibfield  {title} {\bibinfo {title} {{Giant enhancement of
  third-harmonic generation in graphene–metal heterostructures}},\ }\href
  {https://doi.org/10.1038/s41565-020-00808-w} {\bibfield  {journal} {\bibinfo
  {journal} {Nature Nanotechnology}\ }\textbf {\bibinfo {volume} {16}},\
  \bibinfo {pages} {318} (\bibinfo {year} {2021})}\BibitemShut {NoStop}%
\bibitem [{\citenamefont {Siaber}\ \emph
  {et~al.}(2019{\natexlab{a}})\citenamefont {Siaber}, \citenamefont {Zonetti},
  \citenamefont {Cunningham},\ and\ \citenamefont {Sydoruk}}]{Siaber2019}%
  \BibitemOpen
  \bibfield  {author} {\bibinfo {author} {\bibfnamefont {S.}~\bibnamefont
  {Siaber}}, \bibinfo {author} {\bibfnamefont {S.}~\bibnamefont {Zonetti}},
  \bibinfo {author} {\bibfnamefont {J.~E.}\ \bibnamefont {Cunningham}},\ and\
  \bibinfo {author} {\bibfnamefont {O.}~\bibnamefont {Sydoruk}},\ }\bibfield
  {title} {\bibinfo {title} {{Terahertz Plasmon Resonances in Two-Dimensional
  Electron Systems: Modeling Approaches}},\ }\href
  {https://doi.org/10.1103/PhysRevApplied.11.064067} {\bibfield  {journal}
  {\bibinfo  {journal} {Physical Review Applied}\ }\textbf {\bibinfo {volume}
  {11}},\ \bibinfo {pages} {064067} (\bibinfo {year}
  {2019}{\natexlab{a}})}\BibitemShut {NoStop}%
\bibitem [{\citenamefont {Sydoruk}\ \emph {et~al.}(2015)\citenamefont
  {Sydoruk}, \citenamefont {Choonee},\ and\ \citenamefont
  {Dyer}}]{Sydoruk_gate_edge}%
  \BibitemOpen
  \bibfield  {author} {\bibinfo {author} {\bibfnamefont {O.}~\bibnamefont
  {Sydoruk}}, \bibinfo {author} {\bibfnamefont {K.}~\bibnamefont {Choonee}},\
  and\ \bibinfo {author} {\bibfnamefont {G.~C.}\ \bibnamefont {Dyer}},\
  }\bibfield  {title} {\bibinfo {title} {{Transmission and Reflection of
  Terahertz Plasmons in Two-Dimensional Plasmonic Devices}},\ }\href
  {https://doi.org/10.1109/TTHZ.2015.2405919} {\bibfield  {journal} {\bibinfo
  {journal} {IEEE Transactions on Terahertz Science and Technology}\ }\textbf
  {\bibinfo {volume} {5}},\ \bibinfo {pages} {486} (\bibinfo {year}
  {2015})}\BibitemShut {NoStop}%
\bibitem [{\citenamefont {Aizin}\ and\ \citenamefont
  {Dyer}(2012)}]{Aizin_finite}%
  \BibitemOpen
  \bibfield  {author} {\bibinfo {author} {\bibfnamefont {G.~R.}\ \bibnamefont
  {Aizin}}\ and\ \bibinfo {author} {\bibfnamefont {G.~C.}\ \bibnamefont
  {Dyer}},\ }\bibfield  {title} {\bibinfo {title} {{Transmission line theory of
  collective plasma excitations in periodic two-dimensional electron systems:
  Finite plasmonic crystals and Tamm states}},\ }\href
  {https://doi.org/10.1103/PhysRevB.86.235316} {\bibfield  {journal} {\bibinfo
  {journal} {Physical Review B}\ }\textbf {\bibinfo {volume} {86}},\ \bibinfo
  {pages} {235316} (\bibinfo {year} {2012})}\BibitemShut {NoStop}%
\bibitem [{\citenamefont {Moiseenko}\ \emph {et~al.}(2025)\citenamefont
  {Moiseenko}, \citenamefont {Svintsov},\ and\ \citenamefont
  {Nikulin}}]{Moiseenko_partly_gated}%
  \BibitemOpen
  \bibfield  {author} {\bibinfo {author} {\bibfnamefont {I.}~\bibnamefont
  {Moiseenko}}, \bibinfo {author} {\bibfnamefont {D.}~\bibnamefont
  {Svintsov}},\ and\ \bibinfo {author} {\bibfnamefont {E.}~\bibnamefont
  {Nikulin}},\ }\bibfield  {title} {\bibinfo {title} {Electromagnetic
  diffraction and bidirectional plasmon launching in partially gated
  two-dimensional systems},\ }\href {https://doi.org/10.1103/9jw8-5gbp}
  {\bibfield  {journal} {\bibinfo  {journal} {Phys. Rev. Appl.}\ }\textbf
  {\bibinfo {volume} {24}},\ \bibinfo {pages} {014059} (\bibinfo {year}
  {2025})}\BibitemShut {NoStop}%
\bibitem [{\citenamefont {Zabolotnykh}\ and\ \citenamefont
  {Volkov}(2019)}]{Zabolotnykh_proximity}%
  \BibitemOpen
  \bibfield  {author} {\bibinfo {author} {\bibfnamefont {A.~A.}\ \bibnamefont
  {Zabolotnykh}}\ and\ \bibinfo {author} {\bibfnamefont {V.~A.}\ \bibnamefont
  {Volkov}},\ }\bibfield  {title} {\bibinfo {title} {{Interaction of gated and
  ungated plasmons in two-dimensional electron systems}},\ }\href
  {https://doi.org/10.1103/PhysRevB.99.165304} {\bibfield  {journal} {\bibinfo
  {journal} {Physical Review B}\ }\textbf {\bibinfo {volume} {99}},\ \bibinfo
  {pages} {165304} (\bibinfo {year} {2019})}\BibitemShut {NoStop}%
\bibitem [{\citenamefont {Mikhailov}(1998)}]{Mikhailov1998}%
  \BibitemOpen
  \bibfield  {author} {\bibinfo {author} {\bibfnamefont {S.~A.}\ \bibnamefont
  {Mikhailov}},\ }\bibfield  {title} {\bibinfo {title} {{Plasma instability and
  amplification of electromagnetic waves in low-dimensional electron
  systems}},\ }\href {https://doi.org/10.1103/PhysRevB.58.1517} {\bibfield
  {journal} {\bibinfo  {journal} {Physical Review B}\ }\textbf {\bibinfo
  {volume} {58}},\ \bibinfo {pages} {1517} (\bibinfo {year}
  {1998})}\BibitemShut {NoStop}%
\bibitem [{\citenamefont {Falkovsky}\ and\ \citenamefont
  {Varlamov}(2007)}]{Falkovsky2007a}%
  \BibitemOpen
  \bibfield  {author} {\bibinfo {author} {\bibfnamefont {L.~A.}\ \bibnamefont
  {Falkovsky}}\ and\ \bibinfo {author} {\bibfnamefont {A.~A.}\ \bibnamefont
  {Varlamov}},\ }\bibfield  {title} {\bibinfo {title} {{Space-time dispersion
  of graphene conductivity}},\ }\href
  {https://doi.org/10.1140/epjb/e2007-00142-3} {\bibfield  {journal} {\bibinfo
  {journal} {European Physical Journal B}\ }\textbf {\bibinfo {volume} {56}},\
  \bibinfo {pages} {281} (\bibinfo {year} {2007})}\BibitemShut {NoStop}%
\bibitem [{\citenamefont {Dahl}\ \emph {et~al.}(2007)\citenamefont {Dahl},
  \citenamefont {Goy},\ and\ \citenamefont {Kotthaus}}]{dahl2007magneto}%
  \BibitemOpen
  \bibfield  {author} {\bibinfo {author} {\bibfnamefont {C.}~\bibnamefont
  {Dahl}}, \bibinfo {author} {\bibfnamefont {P.}~\bibnamefont {Goy}},\ and\
  \bibinfo {author} {\bibfnamefont {J.~P.}\ \bibnamefont {Kotthaus}},\
  }\bibfield  {title} {\bibinfo {title} {Magneto-optical millimeter-wave
  spectroscopy},\ }in\ \href
  {https://link.springer.com/chapter/10.1007/BFb0103423} {\emph {\bibinfo
  {booktitle} {Millimeter and Submillimeter Wave Spectroscopy of Solids}}}\
  (\bibinfo  {publisher} {Springer},\ \bibinfo {year} {2007})\ pp.\ \bibinfo
  {pages} {221--282}\BibitemShut {NoStop}%
\bibitem [{\citenamefont {Noble}(1958)}]{Noble1958MethodsBO}%
  \BibitemOpen
  \bibfield  {author} {\bibinfo {author} {\bibfnamefont {B.}~\bibnamefont
  {Noble}},\ }\href@noop {} {\emph {\bibinfo {title} {Methods Based on the
  {W}iener-{H}opf Technique for the Solution of Partial Differential
  Equations}}}\ (\bibinfo  {publisher} {Pergamon Press},\ \bibinfo {year}
  {1958})\BibitemShut {NoStop}%
\bibitem [{\citenamefont {Kay}(1959)}]{Kay_reactance_discontinuity}%
  \BibitemOpen
  \bibfield  {author} {\bibinfo {author} {\bibfnamefont {A.}~\bibnamefont
  {Kay}},\ }\bibfield  {title} {\bibinfo {title} {{Scattering of a surface wave
  by a discontinuity in reactance}},\ }\href
  {https://doi.org/10.1109/TAP.1959.1144635} {\bibfield  {journal} {\bibinfo
  {journal} {IRE Transactions on Antennas and Propagation}\ }\textbf {\bibinfo
  {volume} {7}},\ \bibinfo {pages} {22} (\bibinfo {year} {1959})}\BibitemShut
  {NoStop}%
\bibitem [{\citenamefont {Siaber}\ \emph
  {et~al.}(2019{\natexlab{b}})\citenamefont {Siaber}, \citenamefont {Zonetti},\
  and\ \citenamefont {Sydoruk}}]{Retardation_loss}%
  \BibitemOpen
  \bibfield  {author} {\bibinfo {author} {\bibfnamefont {S.}~\bibnamefont
  {Siaber}}, \bibinfo {author} {\bibfnamefont {S.}~\bibnamefont {Zonetti}},\
  and\ \bibinfo {author} {\bibfnamefont {O.}~\bibnamefont {Sydoruk}},\
  }\bibfield  {title} {\bibinfo {title} {Junctions between two-dimensional
  plasmonic waveguides in the presence of retardation},\ }\href
  {https://doi.org/10.1088/2040-8986/ab4056} {\bibfield  {journal} {\bibinfo
  {journal} {Journal of Optics}\ }\textbf {\bibinfo {volume} {21}},\ \bibinfo
  {pages} {105002} (\bibinfo {year} {2019}{\natexlab{b}})}\BibitemShut
  {NoStop}%
\bibitem [{\citenamefont {Ruan}\ and\ \citenamefont {Fan}(2010)}]{Ruan2010}%
  \BibitemOpen
  \bibfield  {author} {\bibinfo {author} {\bibfnamefont {Z.}~\bibnamefont
  {Ruan}}\ and\ \bibinfo {author} {\bibfnamefont {S.}~\bibnamefont {Fan}},\
  }\bibfield  {title} {\bibinfo {title} {{Superscattering of Light from
  Subwavelength Nanostructures}},\ }\href
  {https://doi.org/10.1103/PhysRevLett.105.013901} {\bibfield  {journal}
  {\bibinfo  {journal} {Physical Review Letters}\ }\textbf {\bibinfo {volume}
  {105}},\ \bibinfo {pages} {013901} (\bibinfo {year} {2010})}\BibitemShut
  {NoStop}%
\bibitem [{\citenamefont {Mylnikov}\ and\ \citenamefont
  {Svintsov}(2022)}]{Mylnikov2022}%
  \BibitemOpen
  \bibfield  {author} {\bibinfo {author} {\bibfnamefont {D.}~\bibnamefont
  {Mylnikov}}\ and\ \bibinfo {author} {\bibfnamefont {D.}~\bibnamefont
  {Svintsov}},\ }\bibfield  {title} {\bibinfo {title} {{Limiting capabilities
  of two-dimensional plasmonics in electromagnetic wave detection}},\ }\href
  {https://doi.org/10.1103/PhysRevApplied.17.064055} {\bibfield  {journal}
  {\bibinfo  {journal} {Physical Review Applied}\ }\textbf {\bibinfo {volume}
  {17}},\ \bibinfo {pages} {064055} (\bibinfo {year} {2022})}\BibitemShut
  {NoStop}%
\bibitem [{\citenamefont {Iranzo}\ \emph {et~al.}(2018)\citenamefont {Iranzo},
  \citenamefont {Nanot}, \citenamefont {Dias}, \citenamefont {Epstein},
  \citenamefont {Peng}, \citenamefont {Efetov}, \citenamefont {Lundeberg},
  \citenamefont {Parret}, \citenamefont {Osmond}, \citenamefont {Hong},
  \citenamefont {Kong}, \citenamefont {Englund}, \citenamefont {Peres},\ and\
  \citenamefont {Koppens}}]{Iranzo2018}%
  \BibitemOpen
  \bibfield  {author} {\bibinfo {author} {\bibfnamefont {D.~A.}\ \bibnamefont
  {Iranzo}}, \bibinfo {author} {\bibfnamefont {S.}~\bibnamefont {Nanot}},
  \bibinfo {author} {\bibfnamefont {E.~J.}\ \bibnamefont {Dias}}, \bibinfo
  {author} {\bibfnamefont {I.}~\bibnamefont {Epstein}}, \bibinfo {author}
  {\bibfnamefont {C.}~\bibnamefont {Peng}}, \bibinfo {author} {\bibfnamefont
  {D.~K.}\ \bibnamefont {Efetov}}, \bibinfo {author} {\bibfnamefont {M.~B.}\
  \bibnamefont {Lundeberg}}, \bibinfo {author} {\bibfnamefont {R.}~\bibnamefont
  {Parret}}, \bibinfo {author} {\bibfnamefont {J.}~\bibnamefont {Osmond}},
  \bibinfo {author} {\bibfnamefont {J.~Y.}\ \bibnamefont {Hong}}, \bibinfo
  {author} {\bibfnamefont {J.}~\bibnamefont {Kong}}, \bibinfo {author}
  {\bibfnamefont {D.~R.}\ \bibnamefont {Englund}}, \bibinfo {author}
  {\bibfnamefont {N.~M.}\ \bibnamefont {Peres}},\ and\ \bibinfo {author}
  {\bibfnamefont {F.~H.}\ \bibnamefont {Koppens}},\ }\bibfield  {title}
  {\bibinfo {title} {{Probing the ultimate plasmon confinement limits with a
  van der Waals heterostructure}},\ }\href
  {https://doi.org/10.1126/science.aar8438} {\bibfield  {journal} {\bibinfo
  {journal} {Science}\ }\textbf {\bibinfo {volume} {360}},\ \bibinfo {pages}
  {291} (\bibinfo {year} {2018})}\BibitemShut {NoStop}%
\bibitem [{\citenamefont {Gorbenko}\ and\ \citenamefont
  {Kachorovskii}(2024)}]{Gorbenko_LateralPC}%
  \BibitemOpen
  \bibfield  {author} {\bibinfo {author} {\bibfnamefont {I.}~\bibnamefont
  {Gorbenko}}\ and\ \bibinfo {author} {\bibfnamefont {V.}~\bibnamefont
  {Kachorovskii}},\ }\bibfield  {title} {\bibinfo {title} {Lateral plasmonic
  crystals: Tunability, dark modes, and weak-to-strong coupling transition},\
  }\href {https://doi.org/10.1103/PhysRevB.110.155406} {\bibfield  {journal}
  {\bibinfo  {journal} {Phys. Rev. B}\ }\textbf {\bibinfo {volume} {110}},\
  \bibinfo {pages} {155406} (\bibinfo {year} {2024})}\BibitemShut {NoStop}%
\bibitem [{\citenamefont {Miranda}\ \emph {et~al.}(2024)\citenamefont
  {Miranda}, \citenamefont {Bludov}, \citenamefont {Asger~Mortensen},\ and\
  \citenamefont {Peres}}]{Miranda_2024}%
  \BibitemOpen
  \bibfield  {author} {\bibinfo {author} {\bibfnamefont {D.~A.}\ \bibnamefont
  {Miranda}}, \bibinfo {author} {\bibfnamefont {Y.~V.}\ \bibnamefont {Bludov}},
  \bibinfo {author} {\bibfnamefont {N.}~\bibnamefont {Asger~Mortensen}},\ and\
  \bibinfo {author} {\bibfnamefont {N.~M.~R.}\ \bibnamefont {Peres}},\
  }\bibfield  {title} {\bibinfo {title} {Topology in a one-dimensional
  plasmonic crystal: the optical approach},\ }\href
  {https://doi.org/10.1088/2040-8986/ad8dee} {\bibfield  {journal} {\bibinfo
  {journal} {Journal of Optics}\ }\textbf {\bibinfo {volume} {26}},\ \bibinfo
  {pages} {125001} (\bibinfo {year} {2024})}\BibitemShut {NoStop}%
\bibitem [{\citenamefont {Allen}\ \emph {et~al.}(1977)\citenamefont {Allen},
  \citenamefont {Tsui},\ and\ \citenamefont {Logan}}]{Allen1977}%
  \BibitemOpen
  \bibfield  {author} {\bibinfo {author} {\bibfnamefont {S.~J.}\ \bibnamefont
  {Allen}}, \bibinfo {author} {\bibfnamefont {D.~C.}\ \bibnamefont {Tsui}},\
  and\ \bibinfo {author} {\bibfnamefont {R.~A.}\ \bibnamefont {Logan}},\
  }\bibfield  {title} {\bibinfo {title} {{Observation of the two-dimensional
  plasmon in silicon inversion layers}},\ }\href
  {https://doi.org/10.1103/PhysRevLett.38.980} {\bibfield  {journal} {\bibinfo
  {journal} {Physical Review Letters}\ }\textbf {\bibinfo {volume} {38}},\
  \bibinfo {pages} {980} (\bibinfo {year} {1977})}\BibitemShut {NoStop}%
\bibitem [{\citenamefont {Theis}(1980)}]{Theis_inversion}%
  \BibitemOpen
  \bibfield  {author} {\bibinfo {author} {\bibfnamefont {T.~N.}\ \bibnamefont
  {Theis}},\ }\bibfield  {title} {\bibinfo {title} {Plasmons in inversion
  layers},\ }\href
  {https://doi.org/https://doi.org/10.1016/0039-6028(80)90533-6} {\bibfield
  {journal} {\bibinfo  {journal} {Surface Science}\ }\textbf {\bibinfo {volume}
  {98}},\ \bibinfo {pages} {515} (\bibinfo {year} {1980})}\BibitemShut
  {NoStop}%
\bibitem [{\citenamefont {Khisameeva}\ \emph
  {et~al.}(2025{\natexlab{a}})\citenamefont {Khisameeva}, \citenamefont
  {Shuvaev}, \citenamefont {Zabolotnykh}, \citenamefont {Astrakhantseva},
  \citenamefont {Khudaiberdiev}, \citenamefont {Pimenov}, \citenamefont
  {Kukushkin},\ and\ \citenamefont {Muravev}}]{Khisameeva_crystal}%
  \BibitemOpen
  \bibfield  {author} {\bibinfo {author} {\bibfnamefont {A.~R.}\ \bibnamefont
  {Khisameeva}}, \bibinfo {author} {\bibfnamefont {A.}~\bibnamefont {Shuvaev}},
  \bibinfo {author} {\bibfnamefont {A.~A.}\ \bibnamefont {Zabolotnykh}},
  \bibinfo {author} {\bibfnamefont {A.~S.}\ \bibnamefont {Astrakhantseva}},
  \bibinfo {author} {\bibfnamefont {D.~A.}\ \bibnamefont {Khudaiberdiev}},
  \bibinfo {author} {\bibfnamefont {A.}~\bibnamefont {Pimenov}}, \bibinfo
  {author} {\bibfnamefont {I.~V.}\ \bibnamefont {Kukushkin}},\ and\ \bibinfo
  {author} {\bibfnamefont {V.~M.}\ \bibnamefont {Muravev}},\ }\bibfield
  {title} {\bibinfo {title} {Spectrum of plasma excitations in a plasmonic
  crystal fabricated in an {A}l{G}a{A}s/{G}a{A}s heterostructure},\ }\href
  {https://doi.org/10.1103/3gnt-w1zj} {\bibfield  {journal} {\bibinfo
  {journal} {Phys. Rev. Res.}\ }\textbf {\bibinfo {volume} {7}},\ \bibinfo
  {pages} {033224} (\bibinfo {year} {2025}{\natexlab{a}})}\BibitemShut
  {NoStop}%
\bibitem [{\citenamefont {Bylinkin}\ \emph {et~al.}(2019)\citenamefont
  {Bylinkin}, \citenamefont {Titova}, \citenamefont {Mikheev}, \citenamefont
  {Zhukova}, \citenamefont {Zhukov}, \citenamefont {Belyanchikov},
  \citenamefont {Kashchenko}, \citenamefont {Miakonkikh},\ and\ \citenamefont
  {Svintsov}}]{Bylinkin_tight_binding}%
  \BibitemOpen
  \bibfield  {author} {\bibinfo {author} {\bibfnamefont {A.}~\bibnamefont
  {Bylinkin}}, \bibinfo {author} {\bibfnamefont {E.}~\bibnamefont {Titova}},
  \bibinfo {author} {\bibfnamefont {V.}~\bibnamefont {Mikheev}}, \bibinfo
  {author} {\bibfnamefont {E.}~\bibnamefont {Zhukova}}, \bibinfo {author}
  {\bibfnamefont {S.}~\bibnamefont {Zhukov}}, \bibinfo {author} {\bibfnamefont
  {M.}~\bibnamefont {Belyanchikov}}, \bibinfo {author} {\bibfnamefont
  {M.}~\bibnamefont {Kashchenko}}, \bibinfo {author} {\bibfnamefont
  {A.}~\bibnamefont {Miakonkikh}},\ and\ \bibinfo {author} {\bibfnamefont
  {D.}~\bibnamefont {Svintsov}},\ }\bibfield  {title} {\bibinfo {title}
  {{Tight-Binding Terahertz Plasmons in Chemical-Vapor-Deposited Graphene}},\
  }\href {https://doi.org/10.1103/PhysRevApplied.11.054017} {\bibfield
  {journal} {\bibinfo  {journal} {Physical Review Applied}\ }\textbf {\bibinfo
  {volume} {11}},\ \bibinfo {pages} {054017} (\bibinfo {year}
  {2019})}\BibitemShut {NoStop}%
\bibitem [{\citenamefont {Sai}\ \emph {et~al.}(2023)\citenamefont {Sai},
  \citenamefont {Korotyeyev}, \citenamefont {Dub}, \citenamefont
  {S{\l}owikowski}, \citenamefont {Filipiak}, \citenamefont {But},
  \citenamefont {Ivonyak}, \citenamefont {Sakowicz}, \citenamefont {Lyaschuk},
  \citenamefont {Kukhtaruk}, \citenamefont {Cywi{\'{n}}ski},\ and\
  \citenamefont {Knap}}]{Sai2023}%
  \BibitemOpen
  \bibfield  {author} {\bibinfo {author} {\bibfnamefont {P.}~\bibnamefont
  {Sai}}, \bibinfo {author} {\bibfnamefont {V.~V.}\ \bibnamefont {Korotyeyev}},
  \bibinfo {author} {\bibfnamefont {M.}~\bibnamefont {Dub}}, \bibinfo {author}
  {\bibfnamefont {M.}~\bibnamefont {S{\l}owikowski}}, \bibinfo {author}
  {\bibfnamefont {M.}~\bibnamefont {Filipiak}}, \bibinfo {author}
  {\bibfnamefont {D.~B.}\ \bibnamefont {But}}, \bibinfo {author} {\bibfnamefont
  {Y.}~\bibnamefont {Ivonyak}}, \bibinfo {author} {\bibfnamefont
  {M.}~\bibnamefont {Sakowicz}}, \bibinfo {author} {\bibfnamefont {Y.~M.}\
  \bibnamefont {Lyaschuk}}, \bibinfo {author} {\bibfnamefont {S.~M.}\
  \bibnamefont {Kukhtaruk}}, \bibinfo {author} {\bibfnamefont {G.}~\bibnamefont
  {Cywi{\'{n}}ski}},\ and\ \bibinfo {author} {\bibfnamefont {W.}~\bibnamefont
  {Knap}},\ }\bibfield  {title} {\bibinfo {title} {{Electrical Tuning of
  Terahertz Plasmonic Crystal Phases}},\ }\href
  {https://doi.org/10.1103/PhysRevX.13.041003} {\bibfield  {journal} {\bibinfo
  {journal} {Physical Review X}\ }\textbf {\bibinfo {volume} {13}},\ \bibinfo
  {pages} {041003} (\bibinfo {year} {2023})}\BibitemShut {NoStop}%
\bibitem [{\citenamefont {Khisameeva}\ \emph
  {et~al.}(2025{\natexlab{b}})\citenamefont {Khisameeva}, \citenamefont
  {Shuvaev}, \citenamefont {Moiseenko}, \citenamefont {Gusikhin}, \citenamefont
  {Astrakhantseva}, \citenamefont {Pimenov}, \citenamefont {Svintsov},
  \citenamefont {Kukushkin},\ and\ \citenamefont {Muravev}}]{Accompanying}%
  \BibitemOpen
  \bibfield  {author} {\bibinfo {author} {\bibfnamefont {A.~R.}\ \bibnamefont
  {Khisameeva}}, \bibinfo {author} {\bibfnamefont {A.}~\bibnamefont {Shuvaev}},
  \bibinfo {author} {\bibfnamefont {I.}~\bibnamefont {Moiseenko}}, \bibinfo
  {author} {\bibfnamefont {P.~A.}\ \bibnamefont {Gusikhin}}, \bibinfo {author}
  {\bibfnamefont {A.~S.}\ \bibnamefont {Astrakhantseva}}, \bibinfo {author}
  {\bibfnamefont {A.}~\bibnamefont {Pimenov}}, \bibinfo {author} {\bibfnamefont
  {D.}~\bibnamefont {Svintsov}}, \bibinfo {author} {\bibfnamefont {I.~V.}\
  \bibnamefont {Kukushkin}},\ and\ \bibinfo {author} {\bibfnamefont {V.~M.}\
  \bibnamefont {Muravev}},\ }\bibfield  {title} {\bibinfo {title} {Discovery of
  slot plasma excitations in a {A}l{G}a{N}/{G}a{N} plasmonic crystal},\ }\href
  {https://arxiv.org/abs/2511.03450} {\bibfield  {journal} {\bibinfo  {journal}
  {arxiv preprint 2511.03450}\ } (\bibinfo {year}
  {2025}{\natexlab{b}})}\BibitemShut {NoStop}%
\end{thebibliography}%

\end{document}